\documentclass[12pt]{article}%

\usepackage{a4wide, amsmath, amsthm, amssymb, latexsym, epsfig}
\usepackage{amsfonts}
\usepackage{graphicx}
\usepackage{subfigure}%
\newcommand{\be}{\begin{equation}}
\newcommand{\bea}{\begin{eqnarray}}
\newcommand{\ee}{\end{equation}}
\newcommand{\eea}{\end{eqnarray}}
\newcommand{\bC}{\mathbb C}
\newcommand{\bP}{\mathbb P}
\newcommand{\bR}{\mathbb R}
\newcommand{\bZ}{\mathbb Z}
\newcommand{\cM}{\mathcal M}
\newcommand{\cW}{\mathcal W}
\newcommand{\cA}{\mathcal A}
\newcommand{\cB}{\mathcal B}
\newcommand{\cL}{\mathcal L}
\newcommand{\cN}{\mathcal N}
\newcommand{\ea}{\mit a}
\newcommand{\cD}{\mathcal D}
\newcommand{\gd}{\delta}
\newcommand{\gD}{\Delta}
\newcommand{\gS}{\Sigma}
\newcommand{\Conv}{\mbox{Conv}}
\newcommand{\ret}{\nonumber \\}
\newcommand{\ga}{\alpha}
\newcommand{\gO}{\Omega}
\newcommand{\matr}[2][rrrrrrrrrrrrrrrrrrrr]
 {\left(
  \begin{array}{#1}
   #2\\
  \end{array}
 \right)}

\input epsf.tex
\begin{document}
\bigskip\begin{titlepage}
\begin{flushright}
UUITP-17/07\\
IHES-P/07/33
\end{flushright}
\vspace{2cm}
\begin{center}
{\Large\bf Deforming, revolving and resolving\\} \vspace{0.7cm}
{\Large New paths in the string theory landscape\\}
\end{center}
\vspace{6mm}
\begin{center}
{\large
Diego Chialva{$^{1,a}$}, Ulf   H.\   Danielsson{$^{2,a}$}, Niklas Johansson{$^{3,a}$}, Magdalena~Larfors{$^{4,a}$} and Marcel Vonk{$^{5,b}$}}\\
\vspace{5mm}
{$^a$}Institutionen f\"or Teoretisk Fysik, Box 803, SE-751 08
Uppsala, Sweden. \\
{$^b$}IH\'{E}S, Le Bois-Marie, 35, route 
de Chartres, F-91440 Bures-sur-Yvette, France.\\
\vspace{5mm}
{\tt
{$^1$}diego.chialva@teorfys.uu.se\\
{$^2$}ulf.danielsson@teorfys.uu.se\\
{$^3$}niklas.johansson@teorfys.uu.se\\
{$^4$}magdalena.larfors@teorfys.uu.se\\
{$^5$}mail@marcelvonk.nl\\
}
\end{center}
\vspace{5mm}
\begin{center}
{\large \bf Abstract}
\end{center}
\noindent
In this paper we investigate the properties of series of vacua in the 
string theory landscape. In particular, we study 
minima to the flux potential in type IIB compactifications on 
the mirror quintic. Using geometric transitions, we embed its 
one-dimensional complex structure moduli space in that of another 
Calabi--Yau with $h^{1,1}=86$ and $h^{2,1}=2$. We then show how 
to construct infinite series of continuously connected minima to the 
mirror quintic potential by moving into this larger moduli space, 
applying its monodromies, and moving back. We provide an example of such
series, and discuss their implications for the string theory landscape.
\vfill
\begin{flushleft}
October 2007
\end{flushleft}
\end{titlepage}\newpage

\section{Introduction}
\label{sec:intro}

\bigskip

Compactifications of string theory on six-dimensional internal manifolds provide four-dimensional low-energy effective field theories that could describe our world. (See e.g. \cite{Candelas:1985en}.) Since these models originate from string theory, there is a consistent way of completing them into theories of quantum gravity. However, there is an enormous number of such compactifications. Even if we restrict the internal manifold to be a (conformally) Calabi--Yau threefold the number of possibilities is huge. Furthermore, compactifying string theory on a specific Calabi--Yau leads to a {\em family} of four-dimensional theories, since the moduli of the manifold are unfixed.\footnote{Often, when we speak of ``a Calabi-Yau threefold'', what we really mean is a family of Calabi-Yau threefolds. These threefolds are topologically equivalent, but their shapes and sizes differ, as specified by their complex structure and K{\"a}hler moduli.} Thus the size and shape of the manifold can fluctuate both over three-dimensional space and time. The latter can lead to dynamical problems, such as a rapid decompactification of the theory. 

One way to fix the moduli of the manifold is to introduce fluxes, that pierce certain non-trivial cycles of the manifold.\footnote{See \cite{Grana:2005jc} for a nice review on flux compactifications.} In type IIB these fluxes will create a potential for the complex structure moduli of the Calabi--Yau \cite{DeWolfe:2002nn}. In order to fix the K\"{a}hler moduli we need to take quantum corrections of the theory into account \cite{Kachru:2003aw, Balasubramanian:2005zx}. These effects trap the moduli in metastable minima of the resulting potential, thus stabilizing the compactification. The result is a metastable four-dimensional effective field theory, also known as a string theory vacuum. Each vacuum corresponds to a particular choice of internal manifold, moduli and fluxes. The large number of such vacua form the string theory landscape \cite{Susskind:2003kw}.\footnote{Compactifications of other string theories also yield vacua in the landscape. To be precise, we should note that introducing branes in the compactifications also yields new vacua.}  

A landscape of string theory vacua has many implications for the four-dimensional physics of our world. Some consequences are universal, such as the non-uniqueness and metastability of universes. Other consequences for four-dimensional physics depend on the topography of the landscape, i.e. the distribution of vacua in parameter space, the height and width of potential barriers between vacua, whether the landscape is smooth or rough etc. This will determine the probability for tunneling between vacua, and thus the life-time and evolution of a universe that is described by one vacuum in the landscape. The topographic properties of the landscape are also important for the selection of a particular vacuum, and the computability problems related to this \cite{Denef:2006ad,Acharya:2006zw,Bryng:1995}. 

Mapping out the entire string theory landscape is an enormously difficult task. The landscape is parametrized by a large number of continuous fields and discrete fluxes, and it is difficult to find a systematic way to describe the potential between vacua. However, we can ask ourselves if we can construct models for (parts of) the landscape, and what we can learn from such models.
Examples of such considerations include \cite{Bousso:2000xa, Ceresole:2006iq,Clifton:2007en,Sarangi:2007jb}. In particular, in \cite{Danielsson:2006xw}, it was found that the potential created by three-fluxes in type IIB compactifications often has series of minima that are connected by continuous paths in complex structure moduli space. The construction of such series was based on the use of monodromy transformations.

More specifically, by moving around singular points in the complex structure moduli space, monodromies transform the three-cycles pierced by fluxes. This is equivalent to changing the flux and keeping the cycles fixed. Thus it is possible to move continuously between different minima of the potential, corresponding to different discrete flux values, and still have full control over the potential. Some explicit examples of such series were computed numerically for the mirror quintic, yielding a complete picture of the barriers between minima in this model landscape. 

However, we cannot resolve the barriers between all minima. Not all flux configurations --- 
and hence not all minima of the potential --- are connected 
by monodromy transformations. This suggests a subdivision of the landscape
into several islands. Only minima on the same island are connected by continuous 
paths in complex structure moduli space. 

In the analysis of \cite{Danielsson:2006xw} some questions remained open. The first concerned the length of the series.  
Although only finite series of continuously connected minima were found, there was no general argument as to why infinite series should not exist.\footnote{Note that we are only studying minima of the potential created by fluxes. When we discuss the length of the series of minima we disregard the fixing of K\"{a}hler moduli, which, in e.g. the KKLT model \cite{Kachru:2003aw}, depends on the magnitude of the fluxes. For the sake of argument, we will also disregard questions about back-reaction of the fluxes on the manifold (see e.g. \cite{Grana:2005jc} and references therein). Even though these effects would probably cut off the series of vacua in the string theory landscape at finite length, we expect the remaining series to be long. Hence there are interesting topographic features, such as many closely spaced vacua, in the string theory landscape if these infinite series of minima exist.} One way of indirectly proving the existence of infinite connected series, relates to an interesting mathematical question. The monodromy transformations of the mirror quintic form a subgroup of $Sp(4,\mathbb{Z})$ so only islands connected by such transformations can possibly be connected by monodromies. Furthermore, there are infinite series connected by $Sp(4,\mathbb{Z})$ transformations \cite{Danielsson:2006xw}. This means that if the index of the monodromy group is finite in $Sp(4,\mathbb{Z})$ --- meaning that the number of islands connected by symplectic transformations are finite --- then there are infinite series connected by monodromy transformations. Unfortunately, it is not known whether this index is finite or not \cite{Chen:2006}.

Another question was whether these islands in the landscape really exist, or if they are only an artifact of modeling too small a part of the landscape. In this paper we introduce a new way of computing potential barriers between string theory vacua. We follow new paths that take us to other parts of the landscape. These paths go via topology changing transitions of the internal manifold.

It is well known that many Calabi--Yau manifolds are connected through geometric transitions \cite{Green:1986ck,Candelas:1989ug}. E.g., as we will discuss below, it is reasonable to assume that the mirror quintic, $\mathcal{M}_{(101,1)}$, is connected to a Calabi--Yau threefold with two complex structure moduli and 86 K\"{a}hler moduli, $\mathcal{M}_{(86,2)}$, through a so-called conifold, or geometric,  transition. Thus, we can view the complex structure moduli space of the mirror quintic as a subspace of the moduli space of the other Calabi--Yau. It is possible to make excursions into this larger space and use its monodromies to find continuous paths between minima on the mirror quintic. 

\begin{figure}[tb]
 \begin{center}
  \includegraphics[height=8cm]{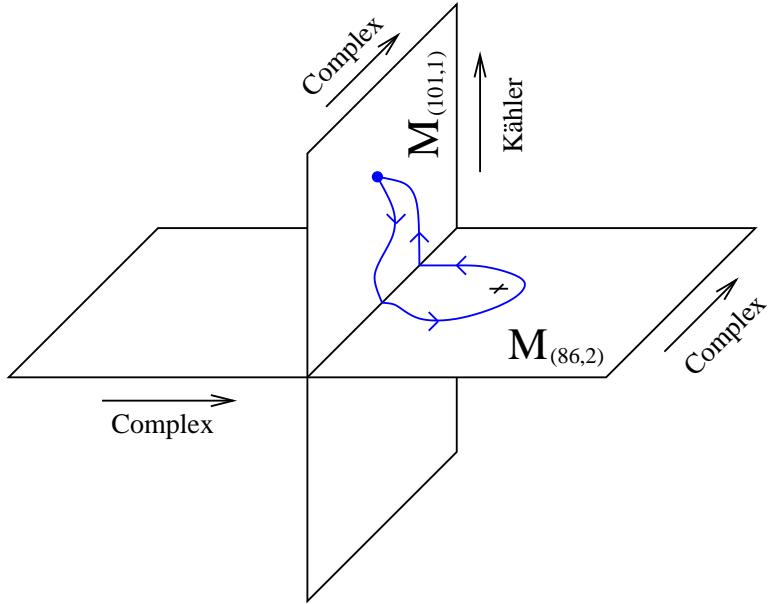}
 \end{center}
 \caption{{\small \sl {The two moduli spaces near the geometric transition. The intersection of the two planes represents the moduli space of the singular manifold, which has 86 remaining K\"ahler moduli and 1 remaining complex structure modulus. Blowing up two-cycles turns the singular manifold into a nonsingular mirror quintic, moving it in its K\"ahler moduli space. Blowing up three-cycles turns the singular manifold into a nonsingular $\cM_{(86,2)}$, moving it in its complex structure moduli space. Note that in particular, we can think of the complex structure moduli space $M_{(101,1)}$ of the mirror quintic as a submanifold of the complex structure moduli space $M_{(86,2)}$ of $\cM_{(86,2)}$. }}}
 \label{fig:modspace}
\end{figure}

To reach the singular point where the conifold transition can happen we need to move in the K\"{a}hler moduli space of the mirror quintic, as shown in figure  \ref{fig:modspace}. It follows that the paths between the minima will go through both the K\"{a}hler moduli space of the mirror quintic and the complex structure moduli space of $\mathcal{M}_{(86,2)}$. In this way, we get new continuous paths between minima in our model landscape. If we have a good description of the potential on both the K\"{a}hler and the complex structure moduli spaces, we get a complete description of the potential barriers between minima. 

This paper will probe deeper into the topography of this model landscape, by extending the complex structure moduli space of the mirror quintic using conifold transitions as described above. To do this, we first compute the geometric data of the Calabi--Yau $\mathcal{M}_{(86,2)}$ using toric geometry (section \ref{sec:toric}). In section \ref{sec:monodromies} we compute the periods and monodromies of $\mathcal{M}_{(86,2)}$ and in section \ref{sec:finding_the_MQ} we show how the complex structure moduli space of $\mathcal{M}_{(101,1)}$ is embedded in the moduli space of $\mathcal{M}_{(86,2)}$. The geometric transition between $\mathcal{M}_{(86,2)}$ and $\mathcal{M}_{(101,1)}$, with and without fluxes, is discussed in section \ref{sec:geomtrans}. In section \ref{sec:infinite_series}, we give an explicit example of a continuously connected series of minima and provide an example of an infinite series. Conventions and a recapitulation of Calabi--Yau geometry are given in appendix \ref{sec:appendix_notation}.

Let us briefly comment on the validity of the techniques used in this paper. The period and monodromy calculations we use are usually applied in the setting of $\cN=2$ compactifications, without fluxes. Here, however, we are interested in the case when flux is turned on. Evading no-go theorems \cite{Maldacena:2000mw, Ivanov:2000fg, Giddings:2001yu} then requires the inclusion of orientifold planes, which reduces the supersymmetry to $\cN=1$. Fortunately, this has little effect for practical calculations: the geometric moduli are kept after the orientifold projection (though they are now part of $\cN=1$ multiplets), and in the large volume limit the terms in the effective $\cN=1$ action correspond directly to the $\cN=2$ terms that are calculated here. Of course, when one leaves this regime there might be $\ga'$-corrections to the potential (coming from cross-cap instantons, for example), but for the purpose of doing monodromy transformations, one can always keep the Calabi-Yau large.

\bigskip

\section{Toric description of $\cM_{(86,2)}$}

\bigskip

\label{sec:toric}
In this paper, we are interested in continuous deformations of the
periods of the mirror quintic. In particular, we want to study how
these periods depend on the coordinates of the complex structure
moduli space. 

The mirror quintic $\mathcal{M}_{(101, 1)}$, with Hodge
numbers $ h^{1,1}=101$ and $h^{2, 1}=1$, is related through mirror
symmetry \cite{Greene:1990ud, Candelas:1990rm} 
to the quintic in $\bP^4$, which has 
$h^{1, 1}=1$ and $h^{2, 1}=101$. The quintic itself is
furthermore known to be connected \cite{Candelas:1989ug,
  Greene:1995hu} 
to a manifold of Hodge
numbers $h^{1, 1}=2$ and $h^{2, 1}=86$ through a geometric
transition.\footnote{When two families of Calabi--Yau manifolds are
related by a geometric transition their moduli spaces are connected
through some loci where the manifolds are singular. The geometric
transition can be 
understood as the shrinking of some three-spheres and the successive
blowing up of the resulting singularities into two-spheres. The
simplest example of this behavior is the transition between resolved
and deformed conifolds. See section \ref{sec:geomtrans} for a more
complete discussion.} It is therefore natural to expect that the
mirror of this process also occurs\footnote{As far as we are aware,
there is no mathematical theorem claiming that the mirror of a
geometric transition is again a geometric transition. However, this
has been suggested to be true in several explicit cases, see e.\
g.\ \cite{Batyrev:1997}. In the present case, we will see explicitly how the
complex structure moduli space of the mirror quintic can be embedded
in that of $\mathcal{M}_{(86, 2)}$.} and that the mirror quintic has a
geometric transition to a Calabi--Yau threefold $\mathcal{M}_{(86, 2)}$
with Hodge 
numbers
\be
 h^{1, 1}=86, \qquad h^{2, 1}=2.
\ee
We will see below that mirror symmetry indeed suggests a natural candidate for this manifold.

We will exploit the above relations between moduli spaces to find very
general continuous deformations of the periods of the mirror
quintic. In particular we will be interested in the behavior of the
periods under the geometric transition between the  mirror quintic and
the manifold $\mathcal{M}_{(86, 2)}$. In order to construct the
manifold and calculate the
periods and their behavior under these deformations, we will use the
tools of {\em toric geometry}.

\subsection{Toric Geometry for Calabi--Yau manifolds}

It is possible to construct Calabi--Yau manifolds as the common zero locus
of a set of polynomials defined on a toric variety. In this subsection we
will briefly review how to obtain these equations and the embedding toric
variety using polytopes of lattice points. The data encoded
in these points are useful to define a suitable set
of coordinates on the complex structure moduli space, and to compute
the periods of the Calabi--Yau threefold (see section \ref{sec:monodromies}).
Moreover, the defining equations for the mirror of a Calabi--Yau can be
represented through another polytope related to the original one,
following a method by V. Batyrev \cite{Batyrev:1994hm}. In this
subsection we will state the general procedure and define the basic concepts in order to
understand it. Readers not well-versed in toric geometry might want to consult
\cite{Kreuzer:2006ax, Cox:2000vi,  Greene:1996cy, Fulton:1993, Hori:2003ic}.  

A $d$-dimensional toric variety $X$ can be constructed as a quotient $X=
\frac{(\bC^*)^n-Z}{G}$, where $G$ is an $m$-dimensional group, and
$d=n-m$. It contains the complex torus
 $(\bC^*)^d$ as an open subset in such a way that the action of the torus on
 itself, given by the group structure, extends to the whole of $X$.

We can either study the toric variety using a set of affine
coordinates $\{t_i,\,\,i=1\ldots d\}$, corresponding to coordinates on
the torus $(\bC^{*})^d$, or with a set of homogeneous ones
$\{x_j, \,\, j=1\ldots n\}$ describing the entire manifold. Using
either of these coordinate sets, we can construct a set of monomials that
form the building blocks for the defining equations of the
Calabi--Yau. In this paper, we will mainly use the affine coordinates. 
 
Let us first discuss the case where the Calabi--Yau manifold is defined
through a single equation.
A convenient way to encode monomials is by points in a
 lattice. Every monomial of the form 
 \be
 \prod_{i=1}^d t_i^{m_i} \equiv t^{\bf m}
 \ee
is represented by a vector of exponents ${\bf m}$ in some lattice
 $M \cong \bZ^d$. More generally, a polytope $\Delta \subset M$
 describes a set of monomials. The vector space given by all the
 linear combinations of such monomials is known as the space of {\em
 Laurent polynomials} associated to the polytope 
$\Delta$. A Laurent polynomial belonging to this space therefore has the form 
\be
 {\bf L}(\{t_i\})= \sum_{l} a_l t^{{\bf m}_l}, \quad {\bf m}_l \in \Delta.
\ee
The set of homogeneous coordinates is defined as follows: given a
polytope $\Delta \subset M$, we can define its dual, or polar, polytope
as 
\be
 \Delta^\circ = \{{{\bf n} \in N | \langle{\bf m}, {\bf n}\rangle \geq -1, \forall {\bf m} \in \Delta}\}.
\ee
Here, $N \cong \bZ^d$ is the lattice dual to $M$. With a simplified notation we 
can write:
 \be
 \langle \Delta^\circ, \Delta \rangle \geq -1.
 \ee

To every vertex
${\bf n}_{j}$ of this polytope we can associate a homogeneous
coordinate $x_j$. The two sets of coordinates are then related by the
equations 
\be 
 \label{affinetohomo}
 t_i = \prod_{j=1}^d x_j^{<{\bf e}_i, {\bf n}_j>}
\ee
where ${\bf e}_i$,  are the basis vectors of $M$. Using this relation,
starting from the Laurent monomials we can define the set of monomials
in the homogeneous coordinates $x_j$ that are invariant under the
group $G$ appearing in the quotient construction of the toric
variety. 

As a side remark, we note that the lattice $N$ is often taken as the
starting point for the construction of a toric variety. The {\em fan}
$\gS$ describing the toric variety is a collection of cones in
$\bR^d \supset N$. The cones of $\gS$ are spanned by the faces
of $\gD^\circ$. In particular, there a one-to-one relation among
the one-dimensional cones, the  
vertices of $\gD^\circ$ and the homogeneous
coordinates. Thus, a polytope $\gD$ contains all the information we
need\footnote{In fact, to make the variety smooth, in certain cases
  one needs a triangulation or a
refinement of $\Delta^\circ$ with additional vertices.
We will not go into these details here.} to construct a
toric variety, which we therefore denote by $X_\gD$. 

We have now collected all the ingredients: the zero locus of the Laurent
polynomials (or their homogenization through (\ref{affinetohomo})) of
$\Delta$ 
defines a Calabi--Yau manifold\footnote{For the conditions on when the
manifold is indeed Calabi--Yau, see e.\
g.\ \cite{Batyrev:1994hm}.} inside the toric variety $X_\Delta$. If $\Delta$ is
{\em reflexive}, meaning roughly that it contains the origin as its
only interior point, the Laurent polynomials that we can instead
obtain from the dual polytope $\Delta^{\circ}$ define its mirror
manifold. This result is the mirror construction of
V.~Batyrev \cite{Batyrev:1994hm}. 

Above, we have described a Calabi--Yau manifold as a hypersurface
 defined by a single polynomial, but we can also study the
 complete\footnote{The completeness condition is a certain regularity
 condition; see \cite{Shafarevich} for details.} intersection of several
 hypersurfaces. In this case, a refined construction is
 necessary \cite{Borisov, Batyrev-1994,Batyrev:1994pg, Klemm:2004km}. 
Once again, the data of the manifold are encoded in a
 lattice polytope, but now this polytope is appropriately partitioned
 into  a set of sub-polytopes\footnote{The relevant sum here and in
 (\ref{nablaintersect}) is the
 Minkowski sum, and the partition must be a so-called {\em nef
 partition} \cite{Batyrev:1994pg, Klemm:2004km}. The term {\em nef} (``numerically effective'') basically
 means that when shifted appropriately, the constituent polytopes
 intersect only in the origin.}: 
 \be
 \Delta = \sum_k \Delta_k
 \ee
Every sub-polytope will encode the data of one of the defining Laurent
 polynomials for the Calabi--Yau. This time, its mirror manifold is
 obtained from a related polytope and a corresponding partition: 
\be \label{nablaintersect}
 \nabla
 = \sum_k \nabla_k. 
\ee
The sub-polytopes $\nabla_k$ are given by 
 \be \label{nablak}
 \nabla_k = \Conv(\{0\}\cup \exists_k), \quad 
 \langle \nabla_k \,,\Delta_{k'}\rangle \geq -\delta_{k, k'}.
 \ee
where, the round brackets indicate the convex hull and the $\exists_k$ are
defined as a partition of the vertices of the dual polytope
$\Delta^\circ$  for
$\Delta$ such that:  
 \be \label{nablakpolar}
 \Delta^\circ = \Conv(\bigcup_k \exists_k).
 \ee

In the next subsection, we will explicitly construct the nef partition
$\{\Delta_k\}$  
 for the manifold
$\mathcal{M}_{(86,2)}$ of our
interest. It is evident from the description above that,
$\nabla$ and $\Delta$ being mirror to each other, we can start the
construction of the mirror pair from an assigned polytope
$\nabla$. In this case we will partition its dual as:
 \be
 \nabla^\circ= \Conv(\bigcup_k E_k),
 \ee
where $\langle \exists_k, E_{k'} \rangle \geq - \delta_{k, k'}$,
and from this, just as above, we will obtain:
 \be
 \Delta_k = \Conv(\{0\}\cup E_k),
 \ee
which we are eventually interested in.

\subsection{Toric data for $\cM_{(86,2)}$}

In order to obtain and list the data for the manifold $\mathcal{M}_{(86,2)}$,
it is convenient to start from those of its mirror, which we denote $\mathcal{W}_{(2, 86)}$. 
Therefore, we obtain the polytope $\Delta$ and its nef partition $\{\Delta_k\}$ relevant for $\mathcal{M}_{(86,2)}$, starting form its mirror $\nabla$ and the nef partition for $\mathcal{W}_{(2, 86)}$. A useful tool in performing the calculations below is the package PALP \cite{PALP}.

A well-known way \cite{Candelas:1989ug, Berglund:1994qk} to describe
the manifold 
$\cW_{(2,86)}$  is 
as a complete intersection of two hypersurfaces given by equations of 
degree $(1,4)$ and $(1,1)$ in $\bP^1 \times \bP^4$. 

These equations can be described by the polytope and the nef
partition\footnote{To describe the equations,we could have taken any 
  shifted version of these polytopes. This particular choice is made in order to obtain
  convenient polynomials in the following subsection.}  
 \be
 \nabla = \nabla_1+\nabla_2
 \ee
where\footnote{Convention: the
  vertices are written as the rows of  the matrix.}
\begin{tiny} 
\[
 \nabla_1 = \Conv \matr{-1 & 0 & -1 & -1 & -1 \\ -1 & 4 & -1 & -1 & -1 \\ -1 & 0 & 3 & -1 & -1 \\ -1 & 0 & -1 & 3 & -1 \\ -1 & 0 & -1 & -1 & 3 \\ 0 & 0 & -1 & -1 & -1 \\ 0 & 4 & -1 & -1 & -1 \\ 0 & 0 & 3 & -1 & -1 \\ 0 & 0 & -1 & 3 & -1 \\ 0 & 0 & -1 & -1 & 3},
 \qquad 
 \nabla_2 = \Conv \matr{0 & -1 & 0 & 0 & 0 \\ 0 & 0 & 0 & 0 & 0 \\ 0 & -1 & 1 & 0 & 0 \\ 0 & -1 & 0 & 1 & 0 \\ 0 & -1 & 0 & 0 & 1 \\ 1 & -1 & 0 & 0 & 0 \\ 1 & 0 & 0 & 0 & 0 \\ 1 & -1 & 1 & 0 & 0 \\ 1 & -1 & 0 & 1 & 0 \\ 1 & -1 & 0 & 0 & 1}. 
\]
\end{tiny}
and therefore:
\begin{tiny}
\[
 \nabla = \Conv \matr{-1 & -1 & -1 & -1 & -1 \\ -1 & 4 & -1 & -1 & -1 \\ -1 & -1 & 4 & -1 & -1 \\ -1 & -1 & -1 & 4 & -1 \\ -1 & -1 & -1 & -1 & 4 \\ 1 & -1 & -1 & -1 & -1 \\ 1 & 4 & -1 & -1 & -1 \\ 1 & -1 & 4 & -1 & -1 \\ 1 & -1 & -1 & 4 & -1 \\ 1 & -1 & -1 & -1 & 4}.
\]
\end{tiny}
$\nabla_1$ and $\nabla_2$ intersect only in the origin  and hence 
form a nef partition of $\nabla$.

The dual polytope of $\nabla$, as explained in the previous
subsection, is given by:
\begin{tiny} 
\[
 \label{eq:dualpol862_2eq}
 \nabla^\circ = \Conv \matr{1 & 0 & 0 & 0 & 0 \\ -1 & 0 & 0 & 0 & 0 \\ 0 & 1 & 0 & 0 & 0 \\ 0 & 0 & 1 & 0 & 0 \\ 0 & 0 & 0 & 1 & 0 \\ 0 & 0 & 0 & 0 & 1 \\ 0 & -1 & -1 & -1 & -1}.
\]
\end{tiny}
Its vertices are in
correspondence with the fan for $\bP^1 \times \bP^4$.\footnote{This fan
  is simply the direct sum of the fans 
for $\bP^1$ and $\bP^4$. It has seven one-dimensional cones: two for
$\bP^1$ and five for $\bP^4$.}

By looking at formulas (\ref{nablaintersect}, \ref{nablak}, \ref{nablakpolar}), we build the nef partition of the polytope related to the  $\mathcal{M}_{(86,2)}$.
First we
subdivide the set $E$ of vertices of
$\nabla^\circ$ in two sets $E = E_1 \cup E_2$: 
\bea
 E_1 & = & \left\{ (1, 0, 0, 0, 0), (0, 0, 1, 0, 0), (0, 0, 0, 1, 0), (0, 0, 0, 0, 1), (0, -1, -1, -1, -1) \right\} \ret
 E_2 & = & \left\{ (-1, 0, 0, 0, 0), (0, 1, 0, 0, 0) \right\}.
\eea
such that $\langle E_k, \nabla_{k'}\rangle \geq -\delta_{k, k'}$. The mirror is now 
constructed by taking the $\gD_k$ to be the convex hulls of $E_k \cup
\{ 0 \}$: 
\begin{tiny} 
\[
 \label{eq:newtonpolytopes862}
 \gD_1 = \Conv \matr{0 & 0 & 0 & 0 & 0 \\ 1 & 0 & 0 & 0 & 0 \\ 0 & 0 & 1 & 0 & 0 \\ 0 & 0 & 0 & 1 & 0 \\ 0 & 0 & 0 & 0 & 1 \\ 0 & -1 & -1 & -1 & -1}, \qquad
 \gD_2 = \Conv \matr{0 & 0 & 0 & 0 & 0 \\ -1 & 0 & 0 & 0 & 0 \\ 0 & 1 & 0 & 0 & 0}.
\]
\end{tiny}
To complete the mirror symmetry circle, one can construct the
Minkowski sum $\gD = \gD_1 + \gD_2$ (which is a polytope with 16
vertices) and its dual $\gD^\circ$ (which is a polytope with 19
vertices), which spans the fan for the manifold in which
$\cM_{(86,2)}$ is a complete intersection. As a consistency check, one
can then show that the mirror of the mirror is the original
manifold. In this paper, we will have no need for the explicit
expressions of the further toric data involved, so we omit them here.

\subsection{Local equations and Mori generators}

As explained previously we can write the defining equation for the
complete intersection in homogeneous or affine coordinates. 
In terms of the affine coordinates, we can conveniently relate
each lattice point in the $\gD_i$  to
a monomial, and we simply read off the
equations: 
\bea
 f_1 \equiv 1 - g_1 & = & 1 - a_1 t_1 - a_2 t_3 - a_3 t_4 - a_4 t_5 -
 a_5 / t_2 t_3 t_4 t_5 \label{eq:laupol1} \\
 f_2 \equiv 1 - g_2 & = & 1 - a_6 / t_1 - a_7 t_2,
 \label{eq:laupol2}
\eea
where we used the convention of scaling the constant term to 1 and
giving all other terms a minus sign. The $a_i$ in these equations are
adjustable constants. It turns out that these local equations are
enough to determine the periods of the holomorphic three-form on
$\cM_{(86,2)}$. We do this calculation in section \ref{sec:monodromies}.  

Varying the $a_i$ changes the complex structure of
our manifold. As can be expected from the fact that the moduli space
has dimension two, not all of the $a_i$ are moduli: only
certain combinations of them are. In fact, we will see in the next
section that it is the constant terms in the products $g_1^n g_2^m$
that determine the moduli. We can expand every power of $g_1$ and $g_2$, 
respectively, in Newton binomials, and therefore these constant
terms will be given by those powers $l_\ell$ of the monomials 
$t^{{\bf m_\ell}}$, ${\bf m}_\ell \in \Delta_1 \cup \Delta_2$,  in
$g_1, g_2$ such that 
 \be
 l_\ell {\bf m}_\ell=0, \quad \ell=1, \ldots |\Delta_1 \cup \Delta_2|.
 \ee
where $|\Delta_1 \cup \Delta_2|$ is the number of points in $\Delta_1
\cup \Delta_2$ (excluding the origin).

It is easy to see that the vectors ${\bf l}$ of the coefficients $\{ l_\ell\}$ of these
relations form a vector space. A convenient choice of basis for
this space is 
\be
 \matr{{\bf l}^{(1)} \\ {\bf l}^{(2)}} = \matr{1 & 0 & 0 & 0 & 0 & 1 & 0 \\0 & 1 & 1 & 1 & 1 & 0 & 1}.
 \label{eq:moricone}
\ee
This basis corresponds to a set of coordinates 
 \be 
 \phi_k = \prod_\ell a_\ell^{l^{(k)}_\ell}
 \ee
 for the  complex structure moduli space, that are particularly useful for describing mirror constructions.
In our case, in fact,  each positive 
linear combination $m l^{(1)} + n l^{(2)}$ corresponds to a power of
$\phi_1^m \phi_2^n$ appearing in some product of $g_1$'s and
$g_2$'s, where
 \be
 \phi_1 = a_1 a_6, \quad \phi_2 = a_2 a_3 a_4 a_5 a_7.
 \label{eq:coord}
 \ee
Thus, the cone\footnote{That is, the linear combinations of ${\bf
l}^{(1, 2)}$ with non-negative coefficients, corresponding to non-negative
powers of the monomials.} generated by ${\bf l}^{(1)}$ and ${\bf
l}^{(2)}$  will play a crucial role in calculating the periods of the holomorphic
three-form. This cone is called the {\em Mori cone},
and the ${\bf l}^{(i)}$ are called the {\em Mori generators}. 

\bigskip

\section{Periods and monodromies of $\mathcal{M}_{(86,2)}$}
\label{sec:monodromies}

\bigskip

In this section we put the machinery of toric geometry to work. Our aim is to calculate
the periods and monodromies of ${\mathcal M}_{(86,2)}$. The 
periods are obtained by finding the solutions to a system
of partial differential equations -- the Picard--Fuchs (PF) equations. Studying the asymptotic
behavior of the periods yields two monodromy transformations of ${\mathcal M}_{(86,2)}$.
 
The methods for computing the periods and monodromies are well-known, but to our
knowledge the periods of this particular manifold have not been explicitly computed before.
(See however \cite{Greene:1996dh}.) A reader familiar with this type of computations can skip the derivations and read the results in equations \eqref{eq:xi0}-\eqref{eq:t_0}.
 
\subsection{The Picard--Fuchs equations}

The periods of a Calabi-Yau 3-fold are the \textquotedblleft holomorphic volumes \textquotedblright of a basis of 3-cycles $C_I$ ($I=1,..,2(h^{2,1}$+1)):
\be
\Pi_I = \oint_{C_I} \Omega(\phi_i),
\ee
where $\Omega (\phi_i)$ is the holomorphic 3-form on the manifold, which depends on the complex structure moduli $\phi_i$.

The periods must satisfy the Picard--Fuchs (PF) equations, as we now explain. Repeated differentiation of $\Omega (\phi_i)$ gives elements in $H^3 = H^{3,0} \oplus H^{2,1} \oplus H^{1,2} \oplus H^{0,3}$. Since $H^3$ has finite dimension, some combinations of derivatives of $\Omega (\phi_i)$ must be exact. Thus $L_k \Omega (\phi_i) = d \eta$, 
where $L_k$ are some differential operators. Integrating over a 3-cycle we find that the periods must fulfill
\be
L_k \Pi_I = \oint_{C_I} L_k \Omega(\phi_i) = \oint_{C_I} d \eta= 0.
\ee
These are the Picard--Fuchs (PF) equations. By solving them, we find the periods.

To derive the PF equations we use a method based on toric geometry, described in \cite{Hosono:1993qy}. We build a set of differential operators -- the generalized hypergeometric Gel'fand--Kapranov--Zelevinski (GKZ) system-- from which, by suitable factorization \cite{Hosono:1993qy,Hosono:1994ax}, we can extract the PF operators. These will be written in terms of the coordinates $\phi_1, \phi_2$ in equation \eqref{eq:coord}, that were defined starting from the coefficients of the Laurent polynomials (\ref{eq:laupol1},\ref{eq:laupol2}) and the generators of the Mori cone \eqref{eq:moricone}.

The differential operators of the GKZ system are also defined from the generators $l^{(k)}$ of the Mori cone \cite{Hosono:1994ax}:
\be
\begin{split}
\mathcal{L}_k &= \prod_{\alpha=1}^r (l_{\alpha}^{(k)}\theta_k)(l_{\alpha}^{(k)}\theta_k-1)...(l_{\alpha}^{(k)}\theta_k-l_{\alpha}^{(k)}+1) \\
        &-\prod_{\beta=1}^s \left( -\sum_{i=1}^k l_{0\beta}^{(i)} \theta_i \right)...\left( -\sum_{i=1}^k l_{0\beta}^{(i)} \theta_i -l_{0\beta}^{(k)} +1 \right)\phi_k,
\end{split}
\ee 
where $\theta_k = \phi_k \frac{\partial}{\partial \phi_k}$, $r$ and $s$ are the dimension and the number of the $l^{(k)}$ respectively, and $l_{0\beta}^{(i)}=-\sum_{\alpha} l_{\alpha}^{(i)}|_{E_{\beta}}$.\footnote{Since we have a nef partition, the sum goes over the $l^{(k)}$ related to the vertices in $E_{\beta}$. See \cite{Hosono:1994ax}.} In our example we obtain:
\be
\begin{split}
\mathcal{L}_1 &= \theta_1^2-(\theta_1+\theta_2)(\theta_1 + 4\theta_2)\phi_1 \\
\mathcal{L}_2 &= \theta_2^5 - (\theta_1+\theta_2)(\theta_1 + 4\theta_2)(\theta_1 + 4\theta_2-1)(\theta_1 + 4\theta_2-2)(\theta_1 + 4\theta_2-3)\phi_2.
\label{eq:GKZ}
\end{split}
\ee
The periods of $\mathcal{M}_{(86,2)}$ are solutions to the equations 
$\cL_k \omega = 0$, but in general there are other solutions as well \cite{Hosono:1993qy}. In fact, the monodromy group does not act irreducibly on the solutions of the GKZ system, as it should do on the periods \cite{Hosono:1994ax}. To get rid of the false solutions, we factorize \cite{Hosono:1993qy,Hosono:1994ax} the differential operators as 
\be
\begin{split}
L_1 &= \mathcal{L}_1\\
p_1 L_2 &= \mathcal{L}_2 - p_3 \mathcal{L}_1,
\end{split}
\ee
where
\be
\begin{split}
L_1 &= \theta_1^2-p_1\phi_1\\
L_2 &= -4\theta_2^3 + 5\theta_1\theta_2^2 + (a\theta_1 + b\theta_2)\theta_1^2 + p_2\phi_1 + p_3\phi_2
\label{eq:PF}
\end{split}
\ee
are the PF operators\footnote{The solutions to the PF equations are the same for all constants $a$ and $b$.}  and 
\be
\begin{split}
p_1&=(\theta_1+\theta_2)(\theta_1 + 4\theta_2) \\
p_2&=-a\theta_1^3-(b+5a)\theta_1^2\theta_2-(5b+4a+5)\theta_1\theta_2^2-(4b+21)\theta_2^3 \\
p_3&= 16(\theta_1 + 4\theta_2-1)(\theta_1 + 4\theta_2-2)(\theta_1 + 4\theta_2-3).
\end{split}
\ee
Note that $L_k \Pi = 0 \implies {\mathcal L}_k \Pi = 0$, but not vice versa. There are six linearly independent solutions to the PF equations, corresponding to the six periods of the $\mathcal{M}_{(86,2)}$.
To compute the periods we need the classical intersection numbers $\kappa_{ijk}$ of $\mathcal{M}_{(86,2)}$. 
These can be read off from the PF operators as was explained in \cite{Hosono:1994ax}. The procedure can be summarized as follows. The numbers are defined as
\be
\kappa_{ijk}=\int J_i \wedge J_j \wedge J_k.
\ee
The $J_k$'s are (1,1)-forms on $\cM_{(86,2)}$ that constitute a basis of the K\"{a}hler cone that is dual to the basis of the Mori cone, $l^{(k)}$. To find the intersection numbers one has to construct the ring of polynomials orthogonal to the ideal generated by $\lim_{\phi \rightarrow 0} L_k(\theta,\phi)$. The top elements of this ring encode the $\kappa_{ijk}$ \cite{Hosono:1994ax}. In our example, we obtain
\be
\kappa_{122}=4 \mbox{,  } \kappa_{222}=5
\label{eq:kappa}
\ee
and all other $\kappa_{ijk}=0$. A non-trivial check of our results is the computation of the Euler characteristic of $\mathcal{M}_{(86,2)}$ \cite{Hosono:1994ax}:
\be
\chi(\mathcal{M}_{(86,2)})=-\frac{1}{3}
\sum_{i,j,k=1}^2 \left( \sum_{\beta=1}^2 l_{0\beta}^{(i)}l_{0\beta}^{(j)}l_{0\beta}^{(k)} + 
\sum_{\alpha=1}^7 l_{\alpha}^{(i)}l_{\alpha}^{(j)}l_{\alpha}^{(k)} \right)\kappa_{ijk}=168,
\label{eq:euler}
\ee
which agrees with $\chi=2(h^{1,1}-h^{2,1})$.

\subsection{Periods and monodromies}

The fundamental period $\omega_0$ is calculated using the Laurent polynomials of the manifold (\ref{eq:laupol1},\ref{eq:laupol2}). We use the formula
\be
\omega_0 = \frac{1}{(2\pi i)^5} \int_{\gamma} \frac{1}{f_1f_2}\frac{dt_1}{t_1}\wedge...\wedge\frac{dt_5}{t_5}
\ee
where the contour $\gamma$ is a product of five circles enclosing $t_i=0$. We need to find the residue of the integrand, which is the constant term in $\frac{1}{f_1f_2}$. This term can be found using the Mori generators. Using $f_i = 1-g_i$ we get
\be
\omega_0 = \frac{1}{(2\pi i)^5} \int_{\gamma} \sum_{m,n=0}^{\infty} g_1^m g_2^n
          \frac{dt_1}{t_1}\wedge...\wedge\frac{dt_5}{t_5},
\ee            
yielding
\be
\omega_0 = \sum_{n_1,n_2} \frac{(n_1+4n_2)!(n_1+n_2)!}{(n_1!)^2(n_2!)^5}\phi_1^{n_1}\phi_2^{n_2},
\label{eq:w0}
\ee
which is convergent near $\phi_i=0$. Here $\phi_k$ are the complex structure moduli defined in equation \eqref{eq:coord}. It is straight-forward to check that $\omega_0$ solves the PF equations \eqref{eq:PF}. 

This period can also be obtained by the Frobenius method. Applying the PF operators to the ansatz
\be
\omega_0 = \sum_{n_1,n_2} c(n_1,n_2)\phi_1^{n_1}\phi_2^{n_2},
\ee
gives recursion relations for $c(n_1,n_2)$ that \eqref{eq:w0} satisfies. Since $\phi_i=0$ is a regular singular point of the PF equations, this is the only power series solution near this point \cite{Hosono:1996gj}. All other solutions contain logarithmic singularities; we obtain two periods with one logarithm, two periods with double logarithms and one period with triple logarithms. Let
\be
\omega(\rho,\phi) = \sum_{n_1,n_2} c(n_1+\rho_1,n_2+\rho_2)\phi_1^{n_1+\rho_1}\phi_2^{n_2+\rho_2}.
\ee
The periods are then given by \cite{Hosono:1994ax}
\be
\begin{split}
\omega_0 &= \omega|_{\rho=0}\\
\omega_1 &= D_1^{(1)}\omega \equiv \frac{1}{2\pi i}\partial_{\rho_1}\omega|_{\rho=0} \\
\omega_2 &= D_2^{(1)}\omega \equiv \frac{1}{2\pi i}\partial_{\rho_2}\omega|_{\rho=0} \\
\omega_3 &= D_1^{(2)}\omega \equiv \frac{1}{2(2\pi i)^2} \kappa_{1jk}\partial_{\rho_j}\partial_{\rho_k}\omega|_{\rho=0} \\
\omega_4 &= D_2^{(2)}\omega \equiv \frac{1}{2(2\pi i)^2} \kappa_{2jk}\partial_{\rho_j}\partial_{\rho_k}\omega|_{\rho=0} \\
\omega_5 &= D^{(3)}\omega \equiv \frac{1}{6(2\pi i)^3} \kappa_{ijk}\partial_{\rho_i}\partial_{\rho_j}\partial_{\rho_k}\omega|_{\rho=0}. \\
\end{split}
\ee
The factors of $2\pi i$ are chosen in order to give integral monodromy matrices. Note that the intersection numbers (\ref{eq:kappa}), related to the leading parts of the PF operators when $\phi_i \rightarrow 0$, determine the
linear combinations of $\partial_{\rho_k}\omega|_{\rho=0}$ that solve the equations.

Naturally, linear combinations of these periods yield other solutions. For our purpose of computing the monodromies of $\mathcal{M}_{(86,2)}$, it is interesting to find a basis of periods which has integral and symplectic monodromy matrices. Furthermore, we would like to find a basis of periods that is symplectic with a canonical symplectic metric, as described in appendix \ref{sec:appendix_notation}.

It was discovered in \cite{Hosono:1993qy,Hosono:1994ax} that such a basis can be constructed, again using toric geometry. How to do this explicitly is clearly described in \cite{Hosono:2000eb}. We start by choosing a basis of $H^{even}(\mathcal{M}_{(86,2)},\mathbb{Q})$: $1, J_k, J_l^{(2)},J^{(3)}$ are forms of degree 0,2,4 and 6 respectively that fulfill
\be
\begin{split}
(1,J^{(3)})&=\int \limits_{\mathcal{M}_{(86,2)}} 1 \wedge J^{(3)} = -1 \\
(J_k,J_j^{(2)})&=\int \limits_{\mathcal{M}_{(86,2)}} J_k \wedge J_j^{(2)} = \delta_{kj}. 
\end{split}
\ee
A canonical symplectic basis of even forms is then found by shifting $J_k \rightarrow J_k^{(1)}=J_k - \frac{c_2 \wedge J_k}{12}$, where $c_2$ is the second Chern class of the manifold \cite{Hosono:2000eb}. This basis is symplectic with respect to the skew symmetric form
\be
q_{\alpha \beta}=\langle \alpha, \beta \rangle = \int \limits_{\mathcal{M}_{(86,2)}} \alpha \wedge (-1)^p \beta \wedge \mbox{Todd} (\mathcal{M}_{(86,2)}),
\ee
where $\alpha$ and $\beta$ are $2q$- and $2p$-forms on $\mathcal{M}_{(86,2)}$, respectively. The Todd class for $\mathcal{M}_{(86,2)}$ is given by
\be
\mbox{Todd}(\mathcal{M}_{(86,2)})=1+c_1+\frac{1}{12}(c_1^2+c_2)+\frac{1}{24}c_1c_2,
\ee
where $c_i$ is the $i^{th}$ Chern class of $\mathcal{M}_{(86,2)}$. Using $c_1=0$, it follows that the symplectic metric for the above basis of even forms is given by
\be
q = \left(
\begin{array}{cccccc}
0&0&0&0&0&1\\
0&0&0&1&0&0\\
0&0&0&0&1&0\\
0&-1&0&0&0&0\\
0&0&-1&0&0&0\\
-1&0&0&0&0&0\\
\end{array}
\right).
\ee
The result of \cite{Hosono:2000eb} is that the period basis
\be
\begin{split}
\xi_0 &= \omega|_{\rho=0}\\
\xi_1 &= D_1^{(1)}\omega \\
\xi_2 &= D_2^{(1)}\omega \\
\xi_3 &= D_1^{(2)}\omega + \frac{1}{2 \pi i}A_{1k}\partial_{\rho_k}\omega|_{\rho=0} \\
\xi_4 &= D_2^{(2)}\omega + \frac{1}{2 \pi i}A_{2k}\partial_{\rho_k}\omega|_{\rho=0} \\
\xi_5 &= D^{(3)}\omega - \frac{1}{2 \pi i}\frac{(c_2,J_k)}{12}\partial_{\rho_k}\omega|_{\rho=0}. 
\label{eq:xi_basis}
\end{split}
\ee
is integral and symplectic with respect to the same $q$. All that is needed to compute the new basis are the  topological numbers $A_{lk}$ and $(c_2,J_k)$, defined as \cite{Hosono:1994ax}
\be
\begin{split}
A_{lk} &= \frac{1}{2} \kappa_{llk} \mbox{ mod } \mathbb{Z} \\
(c_2,J_k) = \int \limits_{\mathcal{M}_{(86,2)}} c_2 \wedge J_k &= \frac{1}{2}\sum_{i,j} \sum_{\alpha} ( l^{(i)}_{0,\alpha} l^{(j)}_{0,\alpha} - l^{(i)}_{\alpha} l^{(j)}_{\alpha} )\kappa_{ijk}.
\end{split}
\ee
Thus, for $\mathcal{M}_{(86,2)}$ we get
\be
\begin{split}
A_{22} &= \frac{1}{2} \mbox{, all other } A_{lk}=0 \\
(c_2,J_1) &= 24 \mbox{ and } (c_2,J_2) = 50.
\end{split}
\ee
Carrying out the differentiations described in \eqref{eq:xi_basis} we then obtain a basis of periods corresponding to a canonical basis of 3-cycles on $\mathcal{M}_{(86,2)}$.

In order to compute the monodromies around the two singular loci $\phi_1=0$ and $\phi_2=0$, we expand $\xi$ near $\phi_i=0$. We have 
\be
\begin{split}
\xi_0 &\sim 1\\
\xi_1 &\sim \frac{1}{2\pi i} \ln \phi_1  \\
\xi_2 &\sim \frac{1}{2\pi i} \ln \phi_2 \\
\xi_3 &\sim -1 + \frac{2}{(2\pi i)^2} \ln^2 \phi_2   \\
\xi_4 &\sim -\frac{25}{12} + \frac{1}{2(2\pi i)} \ln \phi_2 + 
\frac{5}{2(2\pi i)^2}\ln^2 \phi_2 +
\frac{4}{(2\pi i)^2}  \ln \phi_1 \ln \phi_2\\
\xi_5 &\sim \frac{21 i \zeta(3)}{\pi^3} -\frac{25}{12(2\pi i)}\ln \phi_2  - \frac{1}{2\pi i} \ln \phi_1
-\frac{1}{6(2\pi i)^3} \left( 5 \ln^3 \phi_2 + 12 \ln^2 \phi_2 \ln \phi_1 \right). \\
\end{split}
\label{eq:xi0}
\ee

The monodromies around $\phi_i = 0$ are easy to read off from the expansion of the periods around that point, equation \eqref{eq:xi0}. We find that the monodromy around $\phi_1=0$ is given by
\be
t_2 = \left(
\begin{array}{cccccc}
1&0&0&0&0&0\\
1&1&0&0&0&0\\
0&0&1&0&0&0\\
0&0&0&1&0&0\\
0&0&4&0&1&0\\
-2&0&0&-1&0&1\\
\end{array}
\right)
\ee
and the monodromy around $\phi_2=0$ is

\be
t_0 = \left(
\begin{array}{cccccc}
1&0&0&0&0&0\\
0&1&0&0&0&0\\
1&0&1&0&0&0\\
2&0&4&1&0&0\\
3&4&5&0&1&0\\
-5&-2&-2&0&-1&1\\
\end{array}
\right).
\label{eq:t_0}
\ee
We denote the monodromy matrices $t_0$ and $t_2$ in order to conform with the  
notation in \cite{Danielsson:2006xw}: see Appendix \ref{sec:appendix_notation}.

\bigskip
\section{Relating ${\mathcal M}_{(86,2)}$ to the mirror quintic}
\label{sec:finding_the_MQ}

\bigskip

We now set out to find the locus in the moduli space of ${\mathcal M}_{(86,2)}$ that
corresponds to the mirror quintic. Expanding the periods calculated in
section \ref{sec:monodromies} close to this locus will enable us to match periods
between the two manifolds. In doing this, it is important to bear in mind that there are periods 
that vanish on the mirror quintic locus. To make the matching unique we must 
make use of the symplectic structure or, equivalently, of the monodromies around the 
locus. 

Specifically, as we explain more thoroughly in section \ref{sec:geomtrans}, 
one three-cycle\footnote{See appendix \ref{sec:appendix_notation} for notations.} 
$A$ of $H_3({\mathcal M}_{(86,2)},\mathbb{Z})$ shrinks as
we approach the mirror quintic. Its dual $B$ becomes a three-chain as the two-spheres are blown up. 
The cycle $B$ transforms nontrivially as the mirror quintic locus is encircled, 
and neither $A$ nor $B$ should intersect the cycles corresponding to the mirror 
quintic periods. 

\subsection{The mirror quintic locus}

Approaching the mirror quintic locus means sending $\phi_1 \to 1$ and $\phi_2 \to 0$ in such a way that $\phi_2(1-\phi_1)^{-4} \to 0$ while $\phi_1 \phi_2(1-\phi_1)^{-5}$ remains finite. To see this, we use the alternative description of ${\mathcal M}_{(86,2)}$ provided in \cite{Greene:1996dh}, where this locus is determined. If we eliminate the variables $t_1$ and $t_2$ in favor of $t_1 t_2$ in the first equation of (\ref{eq:laupol1}) using \eqref{eq:laupol2}, and make the change of variables and coefficients
\begin{equation}
\begin{array}{llll}
u_1 = t_3, \hskip1cm & u_2 = t_4, \hskip1cm& u_3=t_5, \hskip1cm& u_4 = t_1 t_2\\
b_0 = 1-a_1 a_6, \hskip1cm& b_1 = -a_2, \hskip1cm&  b_2 = -a_3, \hskip1cm&  b_3 = -a_4,\\
b_4 = -a_1 a_7, \hskip1cm& b_5=-a_5 a_6, \hskip1cm& b_6=-a_5 a_7&
\end{array}
\end{equation}
we obtain
\be
b_0 + b_1 u_1 + b_2u_2 + b_3 u_3 + b_4 u_4 + \frac{b_5}{u_1 u_2 u_3 u_4} + \frac{b_6}{u_1 u_2 u_3} = 0.
\ee
This is, up to notation, identical to equation (5.13) of \cite{Greene:1996dh}. The mirror quintic locus is $b_6 = 0$. Natural coordinates on moduli space are, from this point of view, 
\be
z_1 = \frac{b_1b_2b_3b_6}{b_0^4} = \frac{\phi_2}{(1-\phi_1)^4}
\ee
\be
z_2 = -\frac{b_1b_2b_3b_4b_5}{b_0^5} = \frac{\phi_1 \phi_2}{(1-\phi_1)^5}.
\ee
In the patch described by these coordinates the mirror quintic locus is $z_1 = 0$.
We now turn to finding expressions for the periods valid close to this locus. 

\subsection{The fundamental period}

Let us begin with the fundamental period for which it is possible to arrive at a very simple expression
in terms of the variables $(z_1,z_2)$. The period is
\be
\omega_0 = (1-\phi_1)\sum_{n_1,n_2}\frac{(n_1+4n_2)!(n_1+n_2)!}{(n_1!)^2 (n_2!)^5}\phi_1^{n_1}\phi_2^{n_2} = 
\ee
\begin{equation} \label{eq:fundamental}
=(1-\phi_1)\sum_{n_2}\frac{(4n_2)!}{(n_2!)^4}{}_2F_1(4n_2+1,n_2+1,1,\phi_1)\phi_2^{n_2},
\end{equation}
where the prefactor $(1-\phi_1)$ has been chosen to make the period finite at the mirror quintic locus.\footnote{Note that it is always possible to rescale $\Omega$, and hence the periods, by a holomorphic function of the moduli.} We also introduced the hypergeometric function ${}_2F_1$.

By expressing $\omega_0$ in the variables $(z_1,z_2)$ we see explicitly
that $\omega_0$ reduces to the fundamental period of the mirror quintic as $z_1 \to 0$.
Using the analytical continuation of ${}_2F_1$ given in 15.3.4	 of \cite{Abramowitz:1972} we have
\be
{}_2F_1(4n_2+1,n_2+1;1;\phi_1) = (1-\phi_1)^{-4n_2-1} \ {}_2F_1(4n_2+1,-n_2;1;\frac{\phi_1}{\phi_1-1})= 
\ee
\be
= (1-\phi_1)^{-4n_2-1} \sum_{m_2=0}^{n_2} 
\frac{(4n_2+m_{2})!n_2!}{(4n_2)!(n_2-m_{2})!(m_{2}!)^2} \left( \frac{\phi_1}{1-\phi_1} \right)^{m_{2}}.
\ee
The expansion of the ${}_2F_1$ terminates since $n_2$ is a nonnegative integer. Therefore we can replace $n_2$ in favor of $m_1=n_2-m_2$ in the sum representing $\omega_0$, and thus obtain
\begin{equation}
\omega_0 =\sum_{m_{1},m_{2}=0}^{\infty}\frac{(4m_{1}+5m_{2})!}{((m_{1}+m_{2})!)^3 m_{1}! (m_{2}!)^2}z_1^{m_{1}} z_2^{m_{2}}. \label{eq:tildeexpansion}
\end{equation}
The expansion (\ref{eq:tildeexpansion}) can be shown to be convergent for at least $|z_i| < 5^{-3}9^{-3}$. Taking $z_1 \to 0$ we recover the fundamental period of the mirror quintic \cite{Candelas:1990rm} with $z_2 = (5\psi)^{-5}$:
\be
\omega_0(z_1 = 0) =\sum_{m_{2}=0}^{\infty}\frac{(5m_{2})!}{(m_{2}!)^5 (5\psi)^{5m_{2}}}.
\ee

\subsection{The other periods}

For the other periods it is difficult to find explicit expansions in the variables $(z_1,z_2)$, so we just study their values at the mirror quintic locus and their monodromies. The strategy is to analytically continue the ${}_2F_1$ from $\phi_1 \sim 0$
to $\phi_1 \sim 1$ using standard identities. To construct the five additional periods we need the following six derivatives
\be
\partial_{\rho_1}\omega, \quad \partial_{\rho_2}\omega,
\quad \partial_{\rho_2}\partial_{\rho_1}\omega, 
\quad \partial^2_{\rho_2}\omega, \quad \partial^2_{\rho_2}\partial_{\rho_1}\omega,
\quad \partial^3_{\rho_2}\omega,
\ee
with 
\be
\omega(\rho,\phi)= (1-\phi_1)\sum_{n_1,n_2}\frac{\Gamma(\tilde{n}_1+4\tilde{n}_2+1)\Gamma(\tilde{n}_1+\tilde{n}_2+1)}{\Gamma^2(\tilde{n}_1+1) \Gamma^5(\tilde{n}_2+1)}\phi_1^{\tilde{n}_1}\phi_2^{\tilde{n}_2}.
\ee
For brevity we defined $\tilde{n}_i = n_i + \rho_i$. Note that $\omega(0,\phi) = \omega_0(\phi)$

Let us start with the three derivatives containing $\partial_{\rho_1}$. Differentiating once,
evaluating at $\rho_1 = 0$ and
using identity 15.3.10 of \cite{Abramowitz:1972} we obtain
\be
\partial_{\rho_1} \omega(\rho_2) = -(1-\phi_1)\sum_{n_2=0} \frac{\Gamma^2(4\tilde{n}_2+1)}{\Gamma(5\tilde{n}_2+2)\Gamma^3(\tilde{n}_2+1)}
{}_2F_1(4\tilde{n}_2+1,\tilde{n}_2+1;5\tilde{n}_2+2;1-\phi_1)\phi_2^{\tilde{n}_2}. 
\ee 
The above equation is an expansion in $(1-\phi_1)$ and $\phi_2$, both of which go to zero as we approach
the mirror quintic. Because of the overall factor $(1-\phi_1)$ there is no constant term, and differentiating with respect to $\rho_2$ will not change that. Thus all the derivatives $\partial_{\rho_1}\omega$, $\partial_{\rho_2}\partial_{\rho_1}\omega$ and $\partial^2_{\rho_2}\partial_{\rho_1}\omega$ vanish at the mirror quintic locus. These functions will, however, have monodromies around this locus. This is because each
$\partial_{\rho_2}$ that acts on $\phi_2^{\tilde{n}_2}$ produces a factor $\ln \phi_2 = 5\ln(z_1)-4\ln(z_1+z_2)$. It turns out that we need only the behavior of $\partial_{\rho_2}\partial_{\rho_1}\omega$. Carrying out the differentiation yields
\be\label{eq:rho1derivatives}
\begin{split}
&\partial_{\rho_1} \omega \sim 0,\\
&\partial_{\rho_2}\partial_{\rho_1}\omega \sim \ln \phi_ 2 \partial_{\rho_1}\omega.
\end{split}
\ee
Here all terms off the mirror quintic locus that do not contain logarithms are ignored.
For the three derivatives containing only $\partial_{\rho_2}$ we can take $\rho_1 = 0$ and identify the $n_1$ sum as a hypergeometric function exactly as in (\ref{eq:fundamental}). Then, using 15.3.6 of \cite{Abramowitz:1972} it is straightforward to obtain.
\be \label{eq:nice_omega_0}
\omega = 
\sum_{n_2}\frac{\Gamma(5\tilde{n}_2+1)}{\Gamma^5(\tilde{n}_2+1)}(z_1+z_2)^{\tilde{n}_2} 
{}_2F_1(-4\tilde{n}_2,-\tilde{n}_2;-5\tilde{n}_2;1-\phi_1)- f(\rho_2)\partial_{\rho_1}\omega,
\ee
where
\be
f(\rho_2)=\frac{\sin (\pi\rho_2) \sin (4\pi \rho_2)}{\pi \sin (5 \pi \rho_2)}.
\ee
Differentiated with respect to $\rho_2$ and evaluated at $\phi_1=1$ and $z_1 = 0$, the first term of (\ref{eq:nice_omega_0}) yields exactly the four mirror quintic periods. As before, the second term and all its derivatives vanish at the mirror quintic locus, but transform nontrivially under transport around it. Taking the derivatives yields 
\be\label{eq:rho2derivatives}
\begin{split}
&\partial_{\rho_2} \omega \sim \partial_{\rho_2} \omega^{MQ},\\
&\partial^{2}_{\rho_2}\omega \sim \partial^2_{\rho_2} \omega^{MQ} -\frac{8}{5} \partial_{\rho_2}\partial_{\rho_1}\omega,\\
&\partial^3_{\rho_2}\omega \sim \partial^{3}_{\rho_2} \omega^{MQ} 
-\frac{12}{5} \partial^{2}_{\rho_2} \partial_{\rho_1} \omega,
\end{split}
\ee
where $\omega^{MQ} = \omega(\rho_1 = 0; z_1 = 0)$. Again, off the locus $z_1 = 0$, only terms that have monodromies around it are kept.
Combining (\ref{eq:rho1derivatives}) and (\ref{eq:rho2derivatives}) we get for the basis $\xi$
\be
\begin{split}
\xi_0 &\sim \omega|_{\rho=0}\\
\xi_1 &\sim \frac{1}{2\pi i} \partial_{\rho_1}\omega \\
\xi_2 &\sim \frac{1}{2\pi i} \partial_{\rho_2} \omega^{MQ}\\
\xi_3 &\sim \frac{2}{(2\pi i)^2}\partial^2_{\rho_2} \omega^{MQ} - 16\frac{\ln z_ 1}{2\pi i} \xi_1\\
\xi_4 &\sim  \frac{5}{2(2\pi i)^2} \partial^2_{\rho_2} \omega^{MQ} + \frac{1}{2(2\pi i)} 
\partial_{\rho_2} \omega^{MQ}\\
\xi_5 &\sim \frac{5}{6(2\pi i)^3}\partial^{3}_{\rho_2} \omega^{MQ} - \frac{25}{6(2\pi i)}\partial_{\rho_2} \omega^{MQ}. \\
\end{split}
\ee
The terms exhibiting monodromies around $z_1 = 0$ cancel in all basis periods except for $\xi_3$. As the mirror quintic locus is encircled
$\xi_3$ transforms as:
\be
\xi_3 \to \xi_3 - 16 \xi_1.
\label{eq:MQ_conifold}
\ee
We will motivate the appearance of the number $16$ in section \ref{sec:geomtrans}.

\subsection{The proper basis}

From the analysis in the previous subsection it is clear that the periods corresponding to the shrinking three-cycles $A$ and its dual $B$ are $\xi_1$ and $\xi_3$, i.e.,
\be
\xi_1 = \oint_{A} \Omega, \qquad \xi_3 = \oint_{B} \Omega.
\ee
The other four $\xi_i$ correspond to integrals over the cycles that remain in the mirror quintic manifold. By using their asymptotic forms as $z_2 \to 0$, it is straightforward to relate them to the basis for the mirror quintic 
periods used in \cite{Danielsson:2006xw}:
\be
\Pi= m \xi = \left(
\begin{array}
[c]{cccccc}%
0 & 0 & 0 & 0 & 0 & 1\\
0 & 0 & -5 & 0 & -1 & 0\\
0 & 0 & 1 & 0 & 0 & 0\\
1 & 0 & 0 & 0 & 0 & 0\\
0 & 1 & 0 & 0 & 0 & 0\\
0 & 0 & 0 & 1 & 0 & 0
\end{array}
\right)  \xi.
\ee
Here $\Pi_1, \ldots, \Pi_4$ correspond to those of \cite{Danielsson:2006xw}, meaning that $\Pi_5$ and $\Pi_6$ are the 
new periods. In the new basis the symplectic metric becomes%
\be
Q = m q m^{T}=\left(
\begin{array}
[c]{cccccc}%
0 & 0 & 0 & -1 & 0 & 0\\
0 & 0 & 1 & 0 & 0 & 0\\
0 & -1 & 0 & 0 & 0 & 0\\
1 & 0 & 0 & 0 & 0 & 0\\
0 & 0 & 0 & 0 & 0 & 1\\
0 & 0 & 0 & 0 & -1 & 0
\end{array}
\right)  ,
\ee
and the monodromy matrices of section \ref{sec:monodromies} become%
\be
T_0 = m t_0m^{-1}=\left(
\begin{array}
[c]{cccccc}%
1 & 1 & 3 & -5 & -2 & 0\\
0 & 1 & -5 & -8 & -4 & 0\\
0 & 0 & 1 & 1 & 0 & 0\\
0 & 0 & 0 & 1 & 0 & 0\\
0 & 0 & 0 & 0 & 1 & 0\\
0 & 0 & 4 & 2 & 0 & 1
\end{array}
\right)
\ee
and%
\be \label{eq:T2}
T_2=mt_2m^{-1} = \left(
\begin{array}
[c]{cccccc}%
1 & 0 & 0 & -2 & 0 & -1\\
0 & 1 & -4 & 0 & 0 & 0\\
0 & 0 & 1 & 0 & 0 & 0\\
0 & 0 & 0 & 1 & 0 & 0\\
0 & 0 & 0 & 1 & 1 & 0\\
0 & 0 & 0 & 0 & 0 & 1
\end{array}
\right)  .
\ee
The upper four by four corner of $T_0$ precisely coincides with the large
complex structure monodromy for the mirror quintic as given in \cite{Danielsson:2006xw}. 

Two additional monodromies are the conifold monodromies: one around the mirror quintic conifold locus, 
and the other around the mirror quintic locus itself. In the $\Pi$-basis, these are
\be
T_1 = \left(
\begin{array}
[c]{cccccc}%
1 & 0 & 0 & 0 & 0 & 0\\
0 & 1 & 0 & 0 & 0 & 0\\
0 & 0 & 1 & 0 & 0 & 0\\
1 & 0 & 0 & 1 & 0 & 0\\
0 & 0 & 0 & 0 & 1 & 0\\
0 & 0 & 0 & 0 & 0 & 1
\end{array}
\right)
\ee
and%
\be
T_{MQ}=\left(
\begin{array}
[c]{cccccc}%
1 & 0 & 0 & 0 & 0 & 0\\
0 & 1 & 0 & 0 & 0 & 0\\
0 & 0 & 1 & 0 & 0 & 0\\
0 & 0 & 0 & 1 & 0 & 0\\
0 & 0 & 0 & 0 & 1 & 0\\
0 & 0 & 0 & 0 & -16 & 1
\end{array}
\right)  .
\ee
This concludes our description of the geometry of $\cM_{(86,2)}$ and how it reduces to the mirror quintic.  

\bigskip
\section{Geometric transitions in flux compactifications}
\bigskip
\label{sec:geomtrans}

We want to take a closer look at geometric
transitions in the cases where fluxes through the relevant cycles are
involved. Some of these cases have been well-studied in the
literature; others are less understood. 

\subsection{Geometric transition: absence of fluxes}
Let us first describe in some detail what happens to the homology of
the manifold in the case where we make a geometric transition from
$\cM_{(86,2)}$ to the mirror quintic $\cM_{(101,1)}$ {\em without} any
fluxes on the relevant cycles. This discussion is exactly the mirror
of the more familiar story for the ordinary quintic \cite{Candelas:1989ug, Greene:1996dh}. 

\begin{figure}[tb]
 \begin{center}
  \includegraphics[height=4cm]{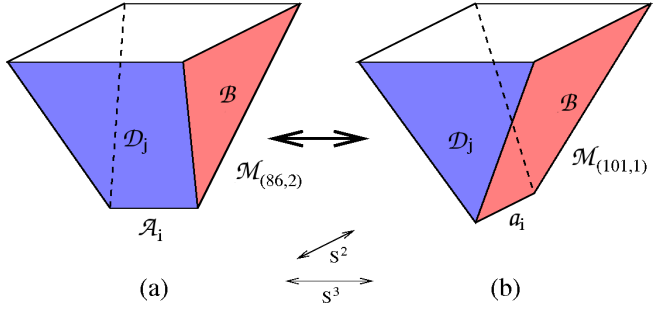}
 \end{center}
 \caption{{\small \sl {The familiar picture of a geometric transition
       on a single node: the $S^3$ denoted by $\cA_i$ shrinks, and the
       $S^2$ denoted by $\ea_i$ blows up. In our example, sixteen of
       those transitions occur simultaneously.}}} 
 \label{fig:transition}
\end{figure}

In the transition, one three-homology class disappears. However, to
obtain the right Hodge numbers we actually need sixteen three-spheres
to shrink. We will denote these three-spheres by $\cA_i, 1 \leq i \leq
16$. Since they are all in the same homology class, there are fifteen
homology relations of the form 
\be
 \cA_1 - \cA_2 = \gd \cD_1 \qquad \cdots \qquad \cA_{15} - \cA_{16} = \gd \cD_{15}
 \label{eq:homology}
\ee
where the $\cD_i$ are four-chains with boundary. For symmetry reasons it
is useful to include a sixteenth four-chain $\cD_{16} = -
\sum_{i=1}^{15} \cD_i$ in the discussion, so that 
$\cA_{16}-\cA_1 = \gd \cD_{16}$. 
Finally, we have also to consider a cycle $\cB$ in the homology
class dual to the class $A$ that the $\{\cA_i\}$'s belong to. This
cycle will intersect the shrinking ones with intersection number 1: 
\be
 \cA_i \cap \cB = 1.
\ee
We have drawn the different cycles and chains in figure \ref{fig:transition}a.

In the transition (see figure \ref{fig:transition}b) 
all sixteen $\cA_i$ shrink to points, and then get blown up into
two-spheres $\ea_i$. These are not homologically independent but form
the boundary of $\cB$: 
\be
 \gd \cB = \sum \ea_i
\ee
Conversely, the $\cD_i$ lose their boundaries and now become closed:
\be
 \gd \cD_i = 0 \qquad 1 \leq i \leq 16
\ee
Of course, we still have that $\sum \cD_i = 0$, so the sixteen cycles
$\cD_i$ actually represent fifteen homology classes. The relations
(\ref{eq:homology}) now turn into intersection relations: 
\bea
 && \cD_1 \cap {\ea_1} = 1, \qquad \cD_1 \cap {\ea_2} = -1 \\
 && \cD_2 \cap {\ea_2} = 1, \qquad \ldots
\eea
and so on.

Summarizing, we see that we lose two dual three-homology classes
$A$ and $B$, corresponding to one complex structure modulus, and
we gain fifteen two-homology classes $a_i$ and fifteen four-homology
classes $D_i$, corresponding to fifteen K\"ahler moduli. 

Finally, the Picard--Lefschetz formula (see for example
\cite{Lefshetz}) says that
 \be \label{PicardLefshtez}
 \cB \to \cB+\sum_{i=1}^{16} (\cB\cap \cA^i) \cA^i =\cB - \sum_{i=1}^{16} \cA^i
 \ee
under transport around the transition locus in moduli space, where the cycles
$\cA^i$ shrink. Thus, the homology classes transform as $B \to B -16A$. 
This explains the appearance of the number 16 in (\ref{eq:MQ_conifold}).

\subsection{Geometric Transition: presence of fluxes}
\label{sec:fluxes}
We now discuss
what happens in a geometric transition if we turn on
fluxes through the relevant cycles $\cA_i$ and $\cB$. There are three
different cases; in some it is clear what happens, in others it is less so.

\vspace{1em}
\noindent
{\em i. Only flux through the $\cA_i$-cycles.}

\begin{figure}[tb]
 \begin{center}
  \includegraphics[height=1.8cm]{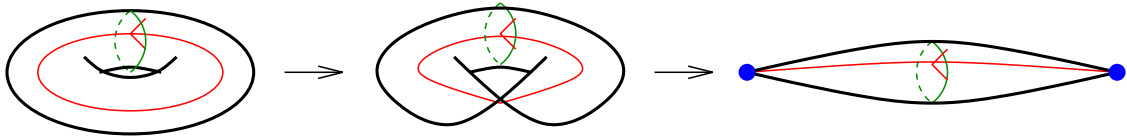}
 \end{center}
 \caption{{\small \sl {The case with flux through the
       shrinking $\cA_i$-cycles. The green line is the cycle, the red line is the
       magnetic flux. As the $\cA_i$-cycles shrink the flux line is cut open. When the two-spheres blow up, 
       five-branes (blue dots) appear,   charging the flux. (Note that a
       zero-dimensional boundary automatically consists of {\em two}
       points; in the conifold case every conifold will of course give
       rise to only {\em one} brane.)}}} 
 \label{fig:fluxesC}
\end{figure}

\vspace{1em}
\noindent
This case has been well-studied in the literature  \cite{Gopakumar:1998ki, Vafa:2000wi}. 
After the geometric transition, we find branes
wrapped on the cycles $\ea_i$. If we start with a single unit of NS-flux
through $\cA_i$, then after the transition we will find a single
NS5-brane wrapping each of the $\ea_i$ and stretching throughout
space-time. These branes are now the sources for the magnetic flux of
the three-form field strength. In a lower-dimensional analogue, we can
think of the magnetic field lines as having been ``cut open'' -- see
figure \ref{fig:fluxesC} for a cartoon. Similarly, if we start with a
single unit of RR-flux through $\cA_i$, we end up with
D5-branes on $\ea_i$. 

We encounter a complication when we want to compute the mirror quintic potential  
for this case, however. When we come back to the mirror quintic
side, in the presence of the five-branes, the closed string moduli are
no longer the only moduli of the system. In particular, as was
explained in \cite{Witten:1997ep, Kachru:2000ih, Kachru:2000an, Aganagic:2000gs}
 (see \cite{Mayr:2002db} for a good review), there will
be some new open string moduli $t_\ga$ describing the the positions of
the branes in $A$. These moduli appear in a new period, which
is the holomorphic volume of the chain $\cB$ with boundary $\sum \ea_i$: 
\be
 \Pi_\cB(t_\ga, z_2) = \int_\cB \gO,
\ee
where the $z_2$ is the closed string modulus. Just like the periods
of the closed cycles, this period contributes to the four-dimensional
superpotential: 
\be
 W_\cB(t_\ga, z_2) = N \Pi_\cB,
\ee
where $N$ is the number of five-branes.

Because of this complication, in our explicit calculations we consider 
the case where we get no five-branes after the transition. It would of 
course be very interesting to extend our calculations to the full case 
including five-branes. 

An interesting possibility of the appearance of five-branes is that we
can now construct domain walls between configurations with and without
five-branes. We will briefly discuss this issue in subsection
\ref{sec:domain_walls}.

\vspace{1em}
\noindent
{\em ii. Only flux through the $\cB$-cycle.}

\vspace{1em}
\noindent
Next, we discuss the case where we only have flux through the dual
cycle $\cB$ of the vanishing cycle, but no flux through the vanishing
cycle itself. In this case, since $\int F_{(3)}\wedge H_{(3)} = 0$, 
the flux does not represent any D3-brane charge, so there is no conservation 
condition that forces the nucleation of e.g. D3-branes. 

Furthermore, as we show in the next subsection, the contribution to the 
scalar potential from this kind of flux vanishes at the transition locus.
Therefore, we expect no problems when going through the geometric transition.\footnote{Indeed, it 
has been argued \cite{Chuang:2005qd} that, after the 
geometric transition, no fluxes or branes 
whatsoever remain. We thank S.~Kachru for useful discussions on this
  case and on the case (iii).}
 
\begin{figure}[tb]
 \begin{center}
  \includegraphics[height=1.8cm]{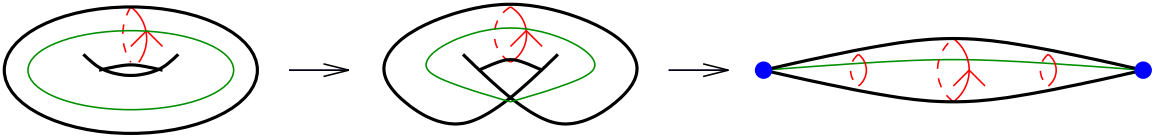}
 \end{center}
 \caption{{\small \sl {The case with flux through the tearing $\cB$-cycle. The
       green line is the cycle, the red line is the flux. After the geometric transition 
       the flux goes through a chain. This flux is possibly charged by one-instantons (blue
       dots).}}} 
 \label{fig:fluxesD}
\end{figure}

It is possible that the flux remains on the other side of the transition, 
sourced by electrically charged D1-instantons or F1-instantons on the blown up two-spheres.
(See \cite{Aganagic:2007py} for a recent discussion on D1-instantons in geometric 
transitions.) A cartoon of this scenario is given in figure \ref{fig:fluxesD}. 
When there is no flux through the $\cA_i$-cycles (that is, no D5-branes are present 
after the transition), these instantons do not contribute to the (super)potential on the resolved side. 

For these reasons, this case is under much more control then
the case (i), and it is the one which we will study in our example
in section \ref{sec:infinite_series}.

\vspace{1em}
\noindent
{\em iii. Flux through the $\cA_i$- and $\cB$-cycles.}

\vspace{1em}
\noindent
This is the most complicated scenario. We will not make any rigorous claims about 
what happens, but let us discuss some possible outcomes. 

First of all, as we mentioned in (i), it is natural to assume that the $\cA_i$-fluxes turn into
five-branes around $\ea_i$. To picture the fate of the fluxes through $\cB$, 
one suggestion comes from the well-known Klebanov--Strassler
setup \cite{Klebanov:2000hb}. In this case, one studies the deformed conifold, which
is a non-compact Calabi--Yau manifold with a compact three-cycle $A$
that shrinks, and a dual non-compact three-cycle $B$. There are $N$
units of RR-flux through the $A$-cycle, and $M$ units on NS-flux
through the $B$-cycle.
The total D3 charge coming from the three-flux background ($\int F_{(3)}\wedge H_{(3)}$) 
and the presence of D3-branes and orientifold
three-planes must be conserved and therefore be equal on the two sides of
the transition. In particular, the above mentioned fluxes piercing the $A$- and
 $B$-cycles give a contribution
$K=MN$ to the total charge. 

After the transition, the deformed conifold becomes a resolved
conifold, which has a new two-cycle $a$ instead of the three-cycle
$A$. One now finds $N$ D5-branes on $a$, as could be expected from the
case (i). Moreover, the contribution to the total charge, previously
coming from the fluxes, now comes from $K=MN$ new D3-branes that fill 
the perpendicular space-time. 
 
In our compact case, the sixteen nodes locally look like conifolds, so
a first possibility is that the same thing happens: if we have
$N$ units of RR-flux through the $\cA_i$ and $M$ units of NS-flux
through $\cB$, we will find $N$ D5-branes on the $\ea_i$ and $K=MN$
space-time filling D3-branes after the transition, so that the charge
conservation condition is satisfied. 

With case (ii) in mind one might instead expect that the fluxes remain,
charged by both five-branes and one-instantons. This situation could either be
stable, or decay into the D3-brane configuration described above.

We may also wonder what happens to fluxes that do not contribute
to the total D3-brane charge. For example, we could have some RR-flux
through $\cA_i$, and some RR-flux through $\cB$ as well. As in case (i)
the flux through $\cA_i$ will result in five-branes. However, no D3-branes will be 
generated, and as in (ii), we might conjecture that after the transition,
either the fluxes through $\cB$ completely disappear or remain charged by one-instantons. 
It would be interesting to investigate these possibilities further. 

Finally, we note that also in this case, as in (i), there will be new
open-string moduli in the theory after the transition. For this
reason, we will not treat this case in our explicit calculations
below. 

\subsection{The scalar potential}

\label{sec:scal_pot}

Fluxes piercing the cycles of $\mathcal{M}_{(86,2)}$ induce a
potential for the two complex structure moduli. It is important to
know the properties of this potential to fully understand the
geometric transition between $\mathcal{M}_{(86,2)}$ and the mirror
quintic. In particular, we need to study the behavior of the potential
as we approach the transition locus $z_1=0$ in the complex
structure moduli space of $\mathcal{M}_{(86,2)}$. (See \cite{Curio:2000sc} for an in-depth study of similar issues.)

We expand the periods near  $z_1=0$ and,
with a convenient notation,
obtain the leading order behavior
\begin{equation}
\Pi(z_1,z_2)\sim\left(
\begin{array}
[c]{c}%
\pi_1(z_2) + {\mathcal O}(z_1)\\
\pi_2(z_2) + {\mathcal O}(z_1)\\
\pi_3(z_2) + {\mathcal O}(z_1)\\
\pi_4(z_2) + {\mathcal O}(z_1)\\
z_1\pi_5(z_2) + {\mathcal O}(z_1^2)\\
\pi_6(z_2)-\frac{16}{2\pi i} z_1\ln(z_1)\pi_5(z_2) + {\mathcal O}(z_1^2\ln(z_1))\\
\end{array}
\right)  .
\end{equation}

The no-scale potential\footnote{See Appendix \ref{sec:appendix_notation} for notations.} is given by 
\be
V\left(  z,\tau\right)  = e^{K}\left(  g^{i\bar{\imath}}D_{i}WD_{\bar{\imath}%
}\overline{W}+g^{\tau\bar{\tau}}D_{\tau}WD_{\bar{\tau}}\overline{W}
\right)
\label{scalarpot}
\ee
where $W$ is the superpotential and $K$ is the K\"{a}hler potential.

Keeping only the leading terms for $z_1 \to 0$, we find
that the potential is given by $V = V_1+V_2$, where:
 \bea
& V_1(z_1, z_2)&  \equiv  e^{K}(  g^{1\bar{1}}D_{1}WD_{\bar{1}}\overline{W})|_{z_1\to 0}              \\
& V_2(z_2)& \equiv e^{K}\left(  g^{2\bar{2}}D_{2}WD_{\bar{2}}\overline{W}
            +g^{\tau\bar{\tau}}D_{\tau}WD_{\bar{\tau}}\overline{W}\right)|_{z_1 \to 0}.
 \eea
Note that terms containing $g^{1\bar{2}}$ and $g^{\bar{1}2}$ are subleading.
This potential has different properties, depending on the presence of
fluxes piercing different cycles.
In particular:
 \begin{itemize}
 \item if there is flux $F_6$, $H_6$ through the shrinking cycle, the
   potential has an infinite spike at the conifold locus. At leading order we
   obtain:
   \be \label{spike}
   V_1 \sim \ln|z_1| \frac{1}{|\tau-\bar{\tau}|} |F_6-\tau H_6|^2 |\pi_5(z_2)|^2.
   \ee
 \item if there is no $F_6$ nor $H_6$, but flux $F_5$, $H_5$ through $B$ we obtain:
   \be
   V_1 \sim \frac{1}{\ln |z_1|}\frac{1}{|\tau-\bar{\tau}|}|F_5-\tau H_5|^2|\pi_5(z_2)|^2
   \ee
   and this part of the potential goes to zero at the conifold
   locus.\footnote{If we minimize the potential with respect to the
     axio-dilaton $\tau$ at the conifold locus, we find the
     result of \cite{Danielsson:2006xw}, i.e. $V_1 \sim
     \frac{1}{\sqrt{K_{z_1 \overline{z}_1}}}$ .}
   This is a general behavior for all compactifications on Calabi--Yau
   threefolds which have conifold singularities, as noted in
   \cite{Danielsson:2006xw}. 
 \end{itemize}
This flux dependence of the potential fits well with the expected
behavior within a geometric transition with different fluxes. 
$V_2$ is the part of the potential that is created by fluxes wrapping 
the mirror quintic cycles. Thus it gives the 
``ordinary'' flux potential for the complex structure modulus $z_2$ on 
the mirror quintic. $V_1$ captures the dependence of the flux through 
the shrinking and torn cycles $A$ and $B$, giving different scenarios.

Let us start with the case having flux through the shrinking cycle $A$ in
$\mathcal{M}_{(86,2)}$, but not through the torn one $B$. As discussed in
section \ref{sec:fluxes}, there will be contributions from five-branes
to the potential for the mirror quintic when we go through
the conifold transition. The behavior of the $\mathcal{M}_{(86,2)}$
potential agrees with this. Indeed, if $F_6$ or $H_6$ are non-zero there are
non-vanishing terms left in $V_1$, even after setting
$z_1=0$. We even find an
infinite spike. When moving away from the conifold point in the
K\"ahler moduli space of the mirror quintic, it is plausible that
the new terms become finite and match the terms coming from a five-brane
contribution. As a side remark, it is tempting to think of the open string
moduli of the branes as a remnant of the complex structure
modulus that one loses. 

On the other hand, if there is zero flux through the $A$ cycle, but
a non-zero one through the $B$ cycle, 
the terms that remain as $z_1 \to 0$ in the $\mathcal{M}_{(86,2)}$ potential
match the terms on the mirror
quintic side. Assuming that we do not generate other terms moving
in the K\"ahler moduli space of the mirror quintic,
there is no need for extra branes after the transition.  

Finally, when we have fluxes through both $A$ and $B$, or only through $A$, we find
terms in $V_1$ 
which remain finite in the limit $z_1
\rightarrow 0$. It is natural to assume that they will also be
present after the transition, and will vary in a
continuous manner as a function of the K\"{a}hler moduli of the mirror
quintic. As mentioned in the previous subsection, it is not clear what the 
physical description of the system after the transition should be, 
but it would be an interesting test of any
proposal for such a description to see if it reproduces these terms in
the appropriate limit. 

\subsection{Flux on the vanishing cycle: domain walls}
\label{sec:domain_walls}

If there is a flux on the vanishing three-cycle, our calculations show
that the potential blows up as we approach the mirror quintic
locus (see equation \eqref{spike}). That is, as the three-spheres contract, the potential
grows. Going through the geometric transition, the vanishing
three-spheres are replaced by vanishing two-spheres, and the flux is
replaced by five-branes wrapping the two-cycles, as discussed
above. 

It is natural to expect that there is a corresponding growth in
the potential if we approach the locus from the side of the mirror
quintic, this time due to the vanishing two-cycles wrapped by
five-branes. We therefore conclude that if all effects from
closed and open string moduli are included, there will be a contribution to the
potential on the mirror quintic side that is inversely
proportional\footnote{It is clear that the potential should grow. For
  arguments as to why it should grow with the inverse volume, see
  \cite{Giddings:2003zw}.} to the volume of the appropriate two-cycles. The
picture 
  we have in mind is that of a spike in the potential, separating the
  phase of the mirror quintic with five-branes and the phase of the
  $\mathcal{M}_{(86,2)}$ with fluxes. 

It is interesting to speculate on the possibility of actually
penetrating through the spike separating the two phases. Classically
this would be possible if the spike is cut off at a finite height by
effects due to stringy and non-perturbative physics. If this is indeed
what happens, there is a possibility of obtaining domain walls
separating regions with and without wrapped five-branes. (See \cite{Aganagic:2007py}
for a recent discussion on transitions between phases with and without branes.)  

Let us start
with the mirror quintic without five-branes, and then move through the
K\"{a}hler moduli space toward the appropriate transition locus that takes us to the
$\mathcal{M}_{(86,2)}$. Without five-branes there is no prominent
barrier that prevents us from doing this. We then encircle an
appropriate locus resulting in a monodromy that generates flux through
the vanishing cycle. The monodromy obtained in (\ref{eq:T2}) is an
example of such a monodromy, as we will see in more detail in the next
section. When we then go back toward the mirror quintic locus, we find a 
high barrier that we need to cross. In contrast to the situation 
we started with, we now have five-branes extending through space-time 
and wrapping internal two-cycles.

We can use this idea to construct domain walls. Let us take one
space-time direction, say the $x$-direction, and put our system in the
phase without five-branes in the region $x \to -\infty$, and in the
phase with five-branes in the region $x \to \infty$. In between these
regions, the system will change in a continuous manner, which will be
exactly described by our path through moduli space. That is,
macroscopically, we will see a domain wall with five-branes ending on
one of its sides. Microscopically, we have a barrier whose profile
is described by the potential as a function of the continuous path
through moduli space.

We can also speculate on the possibility of tunneling from one side to the other. This would be relevant for tunneling between minima connected by
paths in moduli space going through the geometric transition. It is interesting to note that a rough estimate of the tunneling amplitude gives
the result%
\be
 e^{-\int \sqrt{g} V \left( \phi \right) } = e^{-\int_{0}^{\phi_{i}} \sqrt{\ln \phi }\ln \phi } \sim  \mathrm{finite}
\ee
even without a regularization of the singular point $\phi =0$. Assuming that there is no dramatic difference on the K\"{a}hler side of the spike we see
that there is good reason to expect a finite tunneling amplitude.

In the next section we return to the more solid ground of the case with no flux on the vanishing cycle. 

\bigskip
\section{Infinite series of minima}
\bigskip
\label{sec:infinite_series}

As discussed in the introduction, monodromy transformations create continuous paths between minima of the flux-induced potential in Calabi--Yau compactifications \cite{Danielsson:2006xw}. The transformations take us between minima corresponding to different fluxes. Here we investigate the length of such series of minima. We would like to understand if infinite series of continuously connected minima are a topographic feature of our model of the string theory landscape.\footnote{As discussed in the introduction, we are only studying minima of the potential created by fluxes. These will be \emph{vacua} in the string theory landscape if the K\"{a}hler moduli are stabilized and the back-reaction of the fluxes on the manifold can be neglected.} 

In \cite{Danielsson:2006xw}, it was found that infinite series of minima do exist in compactifications on the mirror quintic, but it was unclear if these series can be connected by monodromy transformations. We now use geometric transitions to reach the moduli space of $\mathcal{M}_{(86,2)}$. In this way, we obtain new monodromies, creating new continuous paths between minima. As we now explain, these new transformations yield infinite series of continuously connected minima.

\subsection{General considerations}

In order to find a series of minima we are interested in monodromies $T$
such that the vector of flux quanta\footnote{Our notation is explained in appendix \ref{sec:appendix_notation}.} transforms as%
\be
F_{n}=F_{0}T^{n}=F_{0}+nF_{L},
\ee
which is achieved using a monodromy matrix of the form%
\be
T=\mathbf{1}+\Theta,
\ee
where $\Theta^{2}=0$ and $F_{0}\Theta=F_{L}$. For simplicity we will impose
\be
H_{n}=H_{L},
\ee
with $H_{L}\Theta=0$. If there is a minimum to the potential induced by the flux vectors $F_{L}$ and $H_{L}$,
it follows that a series of fluxes $F_{n}$ and $H_{L}$ provide us with an
infinite series of minima that asymptotes to the minimum given by $F_{L}$
and $H_{L}$ \cite{Danielsson:2006xw}.

As is well-known, the total charge on a compact manifold $\mathcal{M}$ must be zero. The three-form-fluxes, $F_{(3)}$ and $H_{(3)}$, yield a three-brane charge, that must be compensated by charges from three-branes and orientifold planes \cite{Giddings:2001yu}:
\be
\int \limits_{\mathcal{M}} F_{(3)} \wedge H_{(3)}  + Q_{D3} - Q_{O3} = 0.
\label{eq:tadpole}
\ee
If the fluxes are changed in a way that alters $\int_{\mathcal{M}} F_{(3)}  \wedge H_{(3)} = F \cdot Q \cdot H \equiv F \wedge H$, the number of three-branes and orientifold planes must adjust in order to keep this tadpole condition satisfied. This requires the use of new physics, e.g. the nucleation of branes.

On the other hand, if we demand that $F_{L}\wedge H_{L}=0$, the series of minima
will have a constant $F_{n}\wedge H_{n}$ leaving the tadpole condition
unchanged. Thus, in this case, there is no need to nucleate branes. This was the situation studied in \cite{Danielsson:2006xw} and is what we will focus on here.

Let us study the properties of the monodromy matrix $T=\mathbf{1}+\Theta$. Assume that the rank of $\Theta$ is $r$. We can then write
\be
\Theta=%
{\displaystyle\sum\limits_{i=1}^{r}}
\Theta^{\left(  i\right)  }%
\ee
where%
\be
\Theta_{kl}^{\left(  i\right)  }=b_{k}^{\left(  i\right)  }a_{l}^{\left(
i\right)  },
\ee
with $\left\{  b^{\left(  i\right)  }\right\}  $ and $\left\{  a^{\left(
i\right)  }\right\}  $ each being a set of $r$ linearly independent vectors.
It follows from $\Theta^{2}=0$ that
\be
b^{\left(  i\right)  }\cdot a^{\left(  j\right)  }=0 \mbox{ } \forall \mbox{ } i,j=1..r,
\label{eq:ab_cond}
\ee
and we see from
\be
F_{0}\Theta=%
{\displaystyle\sum\limits_{i=1}^{r}}
\left(  F_{0}\cdot b^{\left(  i\right)  }\right)  a^{\left(  i\right)  }%
=F_{L},
\ee
that $F_{L}$ necessarily lies is in the $r$-dimensional subspace spanned by
$\left\{  a^{\left(  i\right)  }\right\}  $. If the dimensionality of the flux
space is $2d,$ it is obvious that we must have $r\leqslant d$.

The difficult part is now to find a monodromy $T=\mathbf{1}+\Theta$ such that
the space $\left\{  a^{\left(  i\right)  }\right\}  $ actually does contain
flux vectors giving rise to a minimum. In \cite{Danielsson:2006xw}, for the case of the mirror
quintic, we were able to find transformations with the right property but unable to show that 
they were monodromies. Let us now examine whether
the extension of the moduli space to the $\mathcal{M}_{(86,2)}$ changes the situation.

\subsection{Series of minima in mirror quintic compactifications}

In section \ref{sec:finding_the_MQ} we generalized the action of the two independent
monodromies on periods for the mirror quintic to periods of the $\mathcal{M}_{(86,2)}$. We also found two new monodromies. Interestingly, our new monodromies are precisely of the form discussed in the
previous subsection. For $T_2=\mathbf{1}+\Theta_2$ we find the rank three matrix%
\be
\Theta_2=\left(
\begin{array}
[c]{cccccc}%
0 & 0 & 0 & -2 & 0 & -1\\
0 & 0 & -4 & 0 & 0 & 0\\
0 & 0 & 0 & 0 & 0 & 0\\
0 & 0 & 0 & 0 & 0 & 0\\
0 & 0 & 0 & 1 & 0 & 0\\
0 & 0 & 0 & 0 & 0 & 0
\end{array}
\right)  ,
\ee
consistent with%
\begin{align}
\begin{array}[c]{ll}%
a^{\left(  1\right)  }  =\left(0, 0, 0,-2, 0,-1\right) 
&  b^{\left(  1\right)  } =\left(1, 0, 0, 0, 0, 0\right)  \\
a^{\left(  2\right)  }  =\left(0, 0,-4, 0, 0, 0\right) 
&   b^{\left(  2\right)  } =\left(0, 1, 0, 0, 0, 0\right)  \\
a^{\left(  3\right)  } =\left(0, 0, 1, 0, 0,0\right) 
&   b^{\left(  3\right)  } =\left(0, 0, 0, 1, 0, 0\right).  
\end{array}
\end{align}
In addition to the conditions \eqref{eq:ab_cond} it is easy to check that also%
\begin{align}
a^{\left(  i\right)  }\wedge a^{\left(  j\right)  } &  =0 \mbox{ } \forall \mbox{ } i,j=0..3\\
b^{\left(  i\right)  }\wedge b^{\left(  j\right)  } &  =0 \mbox{ } \forall \mbox{ } i,j=0..3.
\end{align}
The first of these statements makes sure that $F_{L}\wedge H_{L}=0$ is automatic if
$F_{L}$ and $H_{L}$ lie in the space spanned by the $a^{\left(  i\right)  }$.

We now need to find $F_{L}\in\left\{  a^{\left(  i\right)  }\right\}$ and $H_{L}\in\left\{  a^{\left(
i\right)  }\right\}$ such that when we restrict the flux vectors to the four
dimensional mirror quintic flux space, there is a minimum of the potential. If
we can achieve this we have found an infinite series of continuously connected minima, even
though we have to make geometric transitions to $\mathcal{M}_{(  86,2)}
$ and back again when we pass from one minimum to the next. 

\begin{figure}[tb]
 \begin{center}
\includegraphics[height=10cm]{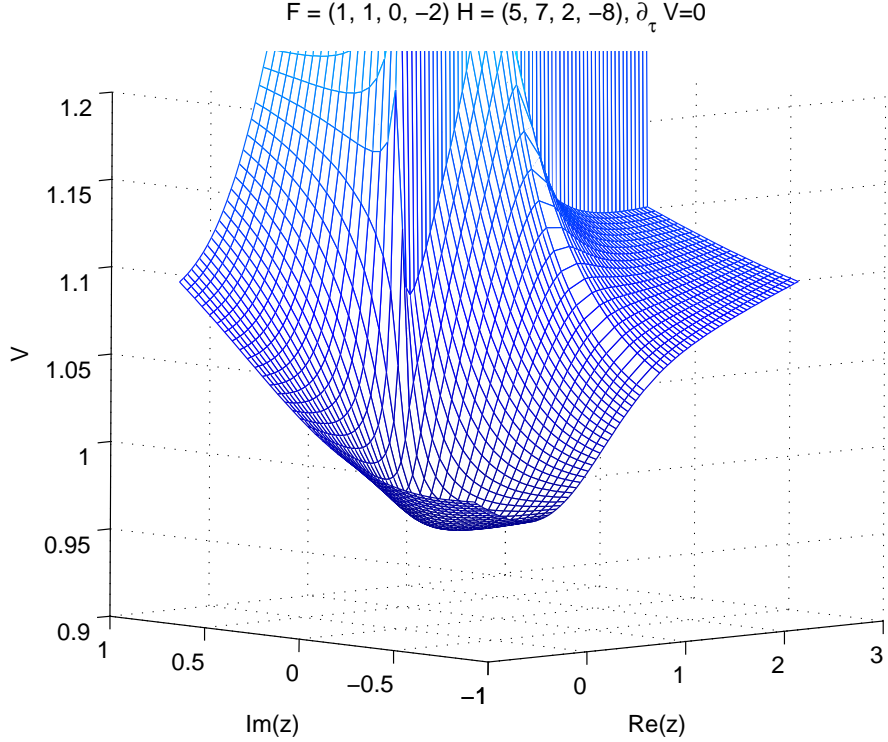}
\end{center}
\caption{{\small \sl {The flux-induced potential $V$ plotted on the complex structure moduli space of the mirror quintic. The plot shows $V(z,\tau(z))$, which is already minimized with respect to the axio-dilaton $\tau$. The complex structure modulus of the mirror quintic is $z=5^{-5}z_2$, where $z_2$ is a complex structure modulus of $\mathcal{M}_{(86,2)}$. The flux vectors are examples of fluxes obtained by applying the monodromy $\tilde{T} = 1 +\tilde{\Theta}$ many times. This minimum proves that there exist infinite series of continuously connected minima.}}} 
 \label{fig:min_plots}
\end{figure}

We have not been able to find any minima using the $T_2=\mathbf{1}+\Theta_2$ above.
However, by the use of the monodromy matrices $T_0$ and $T_1,$
we can rotate the space spanned by $\left\{  a^{\left(  i\right)  }\right\}  $
in such a way that minima are possible to find. A particular example is provided
through the conjugation%
\begin{align}
\tilde{\Theta}& =T_2^{-1}T_1T_0T_1\Theta_2 T_1^{-1}T_0^{-1}T_1^{-1}T_2 \\
\\
& =\left(
\begin{array}
[c]{cccccc}%
2 & 10 & -4 & 4 & 20 & 6\\
0 & 4 & -4 & 4 & 8 & 4\\
-4 & -2 & -4 & 10 & -4 & 3\\
2 & 4 & 0 & -2 & 8 & 1\\
1 & -3 & 4 & -6 & -6 & -4\\
-8 & -4 & -8 & 20 & -8 & 6
\end{array}
\right)
\end{align}

It is straightforward to see that any $F_{L}$ obtained  from $\tilde{\Theta}$ must be of the form%
\be
F_{L}=\left(
\begin{array}
[c]{cccccc}%
f_{1}, & f_{2}, & f_{3}, & -3f_{1}+f_{2}, & 2f_{2}, & \frac{-f_{1}+f_{2}-f_{3}}{2}%
\end{array}
\right)  ,
\label{eq:F_L}
\ee
and correspondingly for $H_{L}$. We can choose the initial flux as
\begin{equation}
F_{0}=\left( 
\begin{array}{cccccc}
-\frac{f_{1}}{2}+\frac{f_{2}}{3}+\frac{f_{3}}{3}, & f_{1}-\frac{f_{2}}{2}-%
\frac{3f_{3}}{4}, & -\frac{f_{1}}{2}+\frac{f_{2}}{6}+\frac{f_{3}}{6}, & 0, & 0,
& 0%
\end{array}%
\right) ,
\end{equation}%
to obtain such a flux vector.

Our numerical investigations show that it is easy to find minima with flux vectors of the form \eqref{eq:F_L}. However, if we want to make sure that we can easily go through the geometric
transition,  there are further restrictions on the flux. The flux that pierces the shrinking cycle induces a  potential barrier at the transition locus, as discussed in subsection \ref{sec:scal_pot}. This corresponds to the generation of five-branes in the geometric
transition. Thus, in order to have a potential that is under full control, the flux through the shrinking cycle must vanish. This implies that the last component of the
flux vector, that is $-f_{1}+f_{2}-f_{3}$, is zero, and correspondingly for $H_L$.   

Even with this restriction it is possible to find minima. One example is shown in figure \ref{fig:min_plots}. The potential in the figure was computed numerically, as
described in appendix \ref{sec:appendix_notation}. The flux configurations, $F_L=(1,1,0,-2)$ and $H_L=(5,7,2,-8)$, correspond to limiting fluxes obtained after many transformations with $\tilde{T} = 
\mathbf{1} +\tilde{\Theta}$. Since acting with $\tilde{T}$ on these fluxes produces a new minimum, we conclude that infinite series of continuously connected minima do exist for our model of the string theory landscape.

\bigskip
\section{Conclusions}
\bigskip
\label{sec:conclusions}

This work was originally triggered by a question left open in a previous
investigation on the topography of the string landscape \cite{Danielsson:2006xw}. 
There, the focus was on determining the existence of
families of string vacua (possibly metastable) connected through
continuous paths in the landscape. The results of that work indicated
that the landscape consists of a set of separate \textquotedblleft
islands\textquotedblright , such that minima on the same island are
connected continuously, but another island could be reached only through
disconnected \textquotedblleft jumps\textquotedblright .  What we have shown
in the present paper is how to enlarge the kind of transformations acting on
the periods and fluxes of a particular theory, and possibly continuously connect islands 
that were not connected within the original setting.

The presence of these new transformations follows from the interconnections among different 
models represented by geometric transitions, and represents a natural characteristic of the landscape. 
In particular, it leads to the discovery of \emph{infinite} series of
continuously connected minima for type IIB string theory in the mirror quintic. An example of
these results can be found in section \ref{sec:infinite_series}. As we have
discussed, we expect effects due in particular to the
stabilization of the K\"{a}hler moduli, which we do not investigate, that
will shorten these series and make them finite. Nevertheless, the existence
of long series of many closely spaced vacua is an interesting topographic
feature of the landscape.

The possibility of reaching all different islands in the landscape through the
use of mirror quintic monodromy transformations, is connected with the unresolved problem
of the finiteness of the index of the monodromy subgroup in $Sp(4,\bZ)$. Our
analysis has shown that there exists continuous transformations that connect
minima in infinite series if we make use of geometric transitions into a
moduli space of larger dimension. The question of whether such
transformations can be found in the original setting remains unanswered. 

The picture of the interconnections between different models,
that we have exploited in order to investigate the series of
minima, represents an interesting result in itself. To our knowledge, the
complete relation between two models related by geometric
transitions has not been made explicit before at the level we do it here.
We were able to find the precise map between the complex structure moduli
spaces of the mirror quintic and the manifold $\mathcal{M}_{(86,2)}$ by analysing 
the behavior of the periods of the latter. This embedding is fully explicit, including
the reduction of the monodromy transformations. 

The analysis was performed neglecting issues regarding K\"{a}hler
moduli and backreaction on the geometry, but took into account, even
if in some cases at a speculative level, the presence of all possible flux
configurations. On the side of the manifold $\mathcal{M}_{(86,2)}$, we are in
full control of the scalar potential of the theory, and we can effectively
study its behavior in certain limits. When we relate it to the potential of the mirror quintic,
though, issues regarding the behavior of the fluxes arise. 

The geometric transition between the two manifolds involves the
shrinking of 16 three-cycles on the $\mathcal{M}_{(86,2)}$ side and the
blowing up of the same number of two-cycles on the mirror quintic one. If no
flux pierces any of these cycles, we have a complete and fully understood
picture of the result of the transformation on the mirror quintic side. If
we instead have fluxes piercing the shrinking cycles and/or the three-cycle
that intersect them, then the picture is more complicated.

Having fluxes only on the shrinking cycles will lead 
to the appearance of D5-branes or NS5-branes wrapping the blown-up
two-cycles on the mirror quintic side \cite{Gopakumar:1998ki, Vafa:2000wi}. 
If the flux pierces only the intersecting cycle, the outcome is less clear. 
Our results show that, at the mirror
quintic locus in the complex structure moduli space of $\mathcal{M}_{(86,2)}$, the contribution
to the (super)potential depending on those fluxes vanishes. Therefore, the transition is not hindered by having such flux.
In both the above cases we have no change in the tadpole condition on the mirror quintic, so no D3-branes need to be nucleated
in the transition. 

Instead if the fluxes pierce both
the shrinking cycles and the intersecting one, the tadpole condition will in
general change with the nucleation of D3-branes as a result. It is not clear
exactly what happens in the present compact case, but it is possible that a
similar nucleation of D3-branes as the one happening in the Klebanov--Strassler
setup will take place.

Different minima imply the presence of domain walls separating them in
space. As analyzed in \cite{Danielsson:2006xw}, the appearance of different islands for
the mirror quintic leads to the conjecture that two different kinds of
domain walls exist. One kind separates fluxes that could be
connected through a monodromy transformation for which a profile depending
on the complex structure moduli can be derived. The other separates
islands that cannot be continuously connected and therefore corresponds
to branes.

In the setup of the present work, we make use of the embedding of the mirror
quintic moduli space into the moduli space of $\cM_{(86,2)}$ in order to derive profiles of a
larger set of domain walls. What happens is that the moduli change when we
cross a domain wall; we leave the mirror quintic moduli
space through a geometric transition to the $\mathcal{M}_{(86,2)}$, and at the
end we come back through another geometric transition to the mirror quintic. 
With no fluxes on the shrinking cycle we have complete control over
the domain wall. When there is a flux through the shrinking cycle the
situation is less clear. If the potential barrier at the geometric
transition generated by the flux through the shrinking cycle is regulated by
stringy corrections, then one can expect to obtain domain walls separating
volumes of space time with different number of five-branes. 

What happens to fluxes through a geometric transition
is a very interesting open question. It leaves open new 
possibilities for finding minima and analysing 
domain walls through the new terms in the scalar potential
generated by the presence of the fluxes. It clearly deserves further
investigation, together with the analysis 
of the dynamics of K\"ahler moduli and backreaction on the
geometry, in order to have a complete picture.

Our results and the techniques employed lead to interesting
possibilities for future research. First of
all, the topographic features of the landscape that we have found
(namely long series of closely spaced vacua) represent a good setting
for chain inflation \cite{Freese:2004vs, Freese:2006fk} (see also \cite{Podolsky:2007vg}) and for the
resonance tunneling \cite{HenryTye:2006tg}. The techniques we have employed, 
and further refinements of them, give hope to obtain a quantitative understanding of
these phenomena. The same techniques can also be used for quantitative study 
of the domain walls between different minima. 

\bigskip
\section*{Acknowledgments}
\bigskip

The work was supported by the Swedish Research Council (VR). The work of D.C. has been supported by the EU Marie Curie Training Site contract: MRTN-CT-2004-512194. We would like to thank Shamit Kachru and Maximilian Kreuzer for useful discussions.

\appendix

\bigskip
\section{Notation and numerics}
\label{sec:appendix_notation}

\bigskip

This appendix explains our notational conventions for the Calabi-Yau geometry and how the 
numerical computations of the scalar potential are performed.

\subsection{Geometry}  

Denote by $\cM$ a Calabi--Yau manifold with complex structure moduli space $M$. The periods of
${\cal M}$ are the \textquotedblleft holomorphic volumes\textquotedblright\ of a basis of 3-cycles:
\begin{equation}
\Pi_{I}=\oint_{C_{I}}\Omega=\oint_{\cM}C_{I}\wedge\Omega.
\end{equation}
Here $\Omega$ is the holomorphic 3-form and $C_{I}$ denotes a basis of
$H_{3}(\cM)$. Note that $C_I$ denotes both the cycles and their Poincar\'{e} duals. The index $I$ runs from $1$ to $2h^{1,2}(\cM)+2\equiv N$. 
The intersection matrix $Q = (Q_{IJ})$ is defined as
\begin{equation}
Q_{IJ}=\oint_{C_{I}}C_{J}=\oint_{\cM}C_{I}\wedge C_{J}.
\end{equation}
We will call $Q_{IJ}$ canonical if the cycles $C_I$, $C_J$ intersect only pairwise, with intersection number $\pm 	1$. 

Denoting the cycles 
corresponding to the $\Pi$-basis of section \ref{sec:finding_the_MQ} by
$C_I$, the cycle that vanishes on the mirror quintic locus is $C_5$ and its dual is $C_6$. We also refer to these cycles by the conventional $A$ and $B$. The cycle that shrinks at the locus intersecting the mirror quintic moduli space in the conifold point of the latter is $C_1$. It is intersected by the cycle $C_4$.

It is customary and convenient to collect the periods into a vector
\begin{equation}
\Pi(z)=\left(
\begin{array}
[c]{c}%
\Pi_{1}(z)\\
\Pi_{2}(z)\\
\vdots\\
\Pi_{N}(z)
\end{array}
\right)  ,
\end{equation}
where $z$ is an $(N/2-1)$-dimensional (complex) coordinate on $M$.

The periods are subject to monodromies:
\begin{equation}
\Pi\rightarrow T \cdot\Pi,
\end{equation}
where $T$ is a matrix that preserves the symplectic structure $Q$.
All possible monodromy matrices constitute a subgroup of $Sp(N,\mathbb{Z})$.

In our example $\cM_{(86,2)}$ we choose a notation for the monodromies that coincides with 
the one used in \cite{Danielsson:2006xw} when reducing to the mirror quintic.\footnote{However, to get a clearer notation, we index the monodromy matrices as $T_i$ instead of $T[i]$ in this paper.} Thus $T_0$ corresponds to
encircling $z_2 = 0$ always staying at the locus $z_1 = 0$.

\subsection{The scalar potential}

Given the periods one can compute the ${\cal N}=1$ scalar potential for given flux quanta.
The standard form of the scalar potential is
\begin{equation}
V\left(  z,\tau\right)  =e^{K}\left(  g^{i\bar{\jmath}}D_{i}WD_{\bar{\jmath}%
}\bar{W}+g^{\tau\bar{\tau}}D_{\tau}WD_{\bar{\tau}}\bar{W}+g^{\rho\bar{\rho}%
}D_{\rho}WD_{\bar{\rho}}\bar{W}-3|W|^{2}\right)  ,\label{superpot}%
\end{equation}
where, as usual, the matrix $g^{A\bar{B}} = (\partial_{A}\partial_{\bar{B}}K)^{-1}$ and 
$D_A = \partial_A + \partial_A K$. In this paper
we focus on the no-scale case, where the contributions of $g^{\rho\bar{\rho}%
}D_{\rho}WD_{\bar{\rho}}\bar{W}$ and $-3|W|^{2}$ cancel:
\begin{equation}\label{eq:noscale}
V\left(  z,\tau\right)  =e^{K}\left(  g^{i\bar{\jmath}}D_{i}WD_{\bar{\jmath}%
}\bar{W}+g^{\tau\bar{\tau}}D_{\tau}WD_{\bar{\tau}}\bar{W} \right). %
\end{equation}
To compute $V$ all that is needed are expressions for the superpotential $W$ and the K\"{a}hler 
potential $K$. We collect the flux quanta in row vectors\footnote{Note that, with this notation, having e.g., $F_1 \neq 0$ 
{\it does not} mean that we have fluxes around $C_1$. Instead it means having fluxes through the intersecting cycle $C_4$. More generally $\int_{C_I} F_{(3)} = -\int_{C_I} \sum_J F_J C_J = \sum_J F_J Q_{JI} = (F \cdot Q)_I \neq F_I$.} $F$ and $H$ according to $F_{(3)}=-\sum_I F_{I}C_{I}=-F\cdot C$ and $H_{(3)}=- \sum_I H_{I}C_{I}=-H\cdot C$.
The superpotential is then given by
\begin{equation}\label{eq:superpotential}
W=\int \limits_{\cM}\Omega\wedge(F_{(3)}-\tau H_{(3)})=F\cdot\Pi-\tau H\cdot\Pi.
\end{equation}
The K\"{a}hler potential is 
\begin{equation}\label{eq:kahlerpotential}
K=-\ln\left(  -i(\tau-\bar{\tau})\right)  +K_{\mathrm{cs}}\left(  z,\bar
{z}\right)  -3\ln\left(  -i(\rho-\bar{\rho})\right).
\end{equation}
$K_{\mathrm{cs}}$ is the K\"{a}hler potential for the complex structure moduli, and is given by
\be
K_{\mathrm{cs}} = -\ln \left( -i\int \limits_{\cM} e^{-4A} \Omega \wedge \bar{\Omega} \right).
\ee
Neglecting warping we have
\begin{equation}\label{eq:kahler}
K_{\mathrm{cs}} = -\ln \left( -i\Pi^{\dagger} \cdot Q^{-1} \cdot \Pi \right).
\end{equation}
Using (\ref{eq:superpotential}), (\ref{eq:kahlerpotential}) and (\ref{eq:kahler})
the scalar potential can be computed numerically once the periods and their 
derivatives are known.

\subsection{Numerics on the mirror quintic}

Let us use the coordinate $z = 5^5 z_2$ on the mirror quintic moduli space.
To find minima of the scalar potential (\ref{eq:noscale}) on the mirror quintic for given fluxes, 
we start by solving for $\tau$ in the equation
\be
\partial_{\tau} V(z,\tau) = 0
\ee
to obtain\footnote{We suppress the dependence of $V$ and $\tau$ on $\bar{z}$ to simplify the notation.} $V(z) = V(z,\tau(z))$. This function has a minimum exactly when $V(z,\tau)$ does. 
We compute the periods $\Pi_{1}(z), \dots, \Pi_{4}(z)$ and their derivatives on a grid in moduli space $M$ using 
the {\it Maple} software package. We use the Meijer-G functions as explained in \cite{Denef:2001xn}. We repeat
the formulas here for the reader's convenience. In $\it Maple$ notation we define
\be
\begin{split}
U^{-}_0(z) &= \mbox{MeijerG}([[4/5, 3/5, 2/5, 1/5], []], [[0], [0, 0, 0]], -z)\\
U^{-}_1(z) &=\frac{1}{2\pi i}\mbox{MeijerG}([[4/5, 3/5, 2/5, 1/5], []], [[0, 0], [0, 0]], z)\\
U^{-}_2(z) &= \frac{1}{(2\pi i)^2}\mbox{MeijerG}([[4/5, 3/5, 2/5, 1/5], []], [[0, 0, 0], [0]], -z)\\
U^{-}_3(z) &=\frac{1}{(2\pi i)^3}\mbox{MeijerG}([[4/5, 3/5, 2/5, 1/5], []], [[0, 0, 0, 0], []], z),
\end{split}
\ee
and 
\be
\begin{split}
U^{+}_0(z) &= U^{-}_0(z)\\
U^{+}_1(z) &=U^{-}_1(z) + U^{-}_0(z)\\
U^{+}_2(z) &= U^{-}_2(z)\\
U^{+}_3(z) &=U^{-}_3(z) + U^{-}_2(z).
\end{split}
\ee
A basis $U_j(z)$ for the mirror quintic periods is then defined by $U_j(z) = U^{-}_j(z)$ for $\mbox{Im(z)}<0$ and
$U_j(z) = U^{+}_j(z)$ for $\mbox{Im(z)}> 0$. The basis $U_j$ is related to the basis $\Pi$ by $\Pi = LU$ with
\be
L = \frac{8i\pi^3}{125}\matr{0 & 5 & 0 & 5 \\ 0 & 3 & -5 & 0\\ 0 & -1 & 0 & 0 \\ 1 & 0 & 0 & 0}.
\ee
The derivatives of the periods $\partial_z \Pi$ needed for computing e.g. $D_z W = \partial_z W + (\partial_z K) W$ 
and $K_{z\bar{z}}$ can also be obtained as Meijer-G functions and calculated in the same way.

Equipped with $\Pi(z)$ and $\partial_z \Pi(z)$ on a grid in moduli space we use {\it Matlab} to efficiently 
compute the potential for a large number of flux vectors. In this way the example of section \ref{sec:infinite_series}
was found.

\end{document}